\documentclass[useAMS,usenatbib]{mn2e}
\usepackage{aas_macro}
\usepackage{epsfig,amsmath,amssymb}
\usepackage{graphicx,rotate,url,mathptmx,lscape,times,color}
\usepackage{color}
\usepackage[varg]{txfonts}

\usepackage{xspace}%% use for \xspace used in some definitions
\usepackage{natbib} %\bibpunct{(}{)}{;}{a}{}{,}% to follow the A&A style

\title[Bi-stability jumps in the winds of massive stars]{Two bi-stability jumps in theoretical wind models for massive stars and the implications for Luminous Blue Variable supernovae}
\author[B. Petrov, J. Vink and G. Gr\"{a}fener]
{Blagovest Petrov\thanks{E-mail: bvp@arm.ac.uk}, Jorick S. Vink and
G\"{o}tz Gr\"{a}fener \\%\thanks{E-mail: daszynska@astro.uni.wroc.pl}\\
Armagh Observatory, College Hill, Armagh, BT61 9DG, Northern Ireland \\ }

\begin{document}

\date{Accepted 2016 February 16. Received 2016 February 16; in original form 2015 April 1}

%\pagerange{\pageref{firstpage}--\pageref{lastpage}} \pubyear{2014}

\maketitle

\label{firstpage}

\begin{abstract}

Luminous Blue Variables  have been suggested to be the direct 
progenitors of supernova types IIb and IIn, with enhanced mass loss prior to explosion. 
However, the mechanism of this mass loss is not yet known.
Here, we investigate the {\bvp qualitative behaviour of theoretical stellar wind 
mass-loss as a function of \Teff
across two bi-stability jumps in blue supergiant regime and also in proximity to the Eddington limit, relevant for LBVs.
To investigate the physical ingredients that play a role in the radiative
acceleration we calculate blue supergiant wind models with the {\sc cmfgen} non-LTE model atmosphere 
code over an effective temperature range  between 30\,000 and 8\,800\,K.}
Although our aim is not to provide new mass-loss rates for BA supergiants, 
we study and confirm the existence of two bi-stability jumps in mass-loss rates predicted by \citet*{vink99}. 
{\bvpb However, they are found to occur at somewhat lower \Teff (20\,000 and 9\,000\,K, respectively) than found previously, which would imply that stars may evolve
towards lower \Teff before strong mass-loss is induced by the bi-stability jumps.
}
When the combined effects of the second bi-stability jump and the proximity to Eddington limit are accounted for, we find a dramatic increase in the mass-loss rate by up to a factor of 30.
Further investigation of both bi-stability jumps is expected to lead to a better understanding of 
discrepancies between empirical modelling and theoretical mass-loss rates reported in the literature,
and to provide key inputs for the evolution of both 
normal AB supergiants and LBVs, as well as their subsequent supernova type II explosions.

\end{abstract}

\begin{keywords}
stars: mass loss --  stars: supergiants -- stars: atmospheres -- stars: winds, outflows
\end{keywords}

%%%%%%%%%%%%%%%%%%%%%%%%%%%%%%%%%%%%%%%%%%%%%%%%
%INTRODUCTION
%%%%%%%%%%%%%%%%%%%%%%%%%%%%%%%%%%%%%%%%%%%%%%%%
\section{Introduction}

Luminous Blue Variables (LBVs) are unstable massive stars in close 
proximity to the Eddington limit \citep{hd94,vink12}. They 
are characterised by strong stellar winds (with mass-loss rates of up to $10^{-3}$\,\msunpyr) 
and large variations in their visual magnitudes ($\Delta V\simeq 1-2$) due 
to variable radii and \Teff in the range $\sim8\,000-30\,000$\,K.
During these ``S Doradus'' excursions LBVs cross the temperature range of two bi-stability jumps (BSJ),
which is expected to lead to winds with variable mass-loss rate and terminal velocity \citep*{vink99,vink02,groh11a}.

Traditionally, stellar evolution models have considered 
LBVs to be in a transitory phase between H-burning O-type stars and
He-burning Wolf-Rayet stars \citep{langer94,maeder00,ekstrom12}. However, this has not been
supported by recent observations which indicate that some LBVs might be
direct progenitors of both supernovae (SNe) type IIb and IIn \citep{kotak06,smith07,trundle08,gal-yam09}. 
More recently, evolutionary calculations of both single stars \citep*{groh13} and 
binary mergers \citep{justham14} have been able to reproduce LBVs as direct 
SN type II progenitors.   

If LBVs are indeed in a direct pre-SN state, 
the fact that they reside in close  
proximity to the Eddington and bi-stability limits, may dramatically affect their mass-loss 
rates prior to explosion. 
Indeed, both SN IIn \citep{kiewe12,ofek13,moriya14} and 
IIb SNe, such as 2013cu 
\citep{gal-yam14,groh14b,grafener16} 
seem to have been subjected to increased mass loss prior to explosion.  
This increased mass loss is oftentimes attributed to ``eruptive'' mass loss \citep{smith15}, although 
a physical mechanism for such eruptive mass loss has not been agreed upon. The 
wave-driven mass-loss mechanism by \citet{shiode14} may be a good candidate, but 
how this would produce an outflow has yet to be determined. 
The most promising scenario for eruptive mass loss has been proposed by \citet{owocki04}, who developed a theory of 'porosity-moderated mass loss'
\citep[see also][]{shaviv98,shaviv00}.
As LBVs find themselves in close proximity to the Eddington limit, we can be 
confident that radiation pressure will play a role in the driving of an outflow \citep{owocki15,vink15}. 
This is even more likely now 
that empirical mass-loss rates in close proximity to the Eddington limit have been found to increase 
steeply \citep{bestenlehner14} -- in agreement with theoretical expectations of
\cite{grafener11,vink11,vink12b}.

Eruptive mass loss has also been proposed as a necessary ingredient in the overall  
mass loss of massive stars during their lives. 
The reason that the stellar wind mass-loss rates during the O-star phase have 
been reduced, as empirical O star mass-loss rates have been down-revised 
due to wind clumping. However, as these reduced rates 
agree reasonably well (within a factor of 2 to 3) with the theoretical rates of \citet*{vink00,vink01} that are commonly used 
in stellar evolution models \citep{martins13}, it is not so clear eruptive mass loss is 
relevant \citep[see also][]{khan15}
. Instead, stellar winds in the OBA supergiant range likely dominate
over those of early O star mass-loss rates \citep{groh14}.

The bi-stability jump \citep{pauldrach90} is observed as a drop in the
  ratio between the terminal and escape velocities
  ($\varv_{\infty}/\varv_{\rm esc}$) of B-type supergiants (\bsgs) by a factor 2,
  when the effective temperature (\Teff) falls 
  below\,$\sim$21\,000 K
  \citep{lamers95,cro06,markova08,garcia14}. \citet{vink99} showed that this was the result of 
  the recombination of the dominant wind driver from \FeIV to \FeIII. They also predicted
  that this drop should be accompanied by a 
   jump in the mass-loss rate
  (\mdot) which may have been confirmed \textit{qualitatively} by  \cite{benaglia07}, but significant inconsistencies between theoretical and 
  empirical mass-loss rates have also remained
  (\citealt{vink00}; \citealt{trundle04}; \citealt{trundle05}; \citealt*{cro06}).
  These discrepancies might be due to 
  the issues of wind
  clumping (\citealt{massa03}; \citealt*{bouret05}; \citealt{puls06}; \citealt*{full06}; \citealt*{davies07}; \citealt*{puls08}), inadequate
  treatment of macro-clumping and porosity effects
  \citep{oskinova07,prinja10,sundq10,sundqvist11,muijres11,surlan12,petrov14}.
  {\bvpb
  In any case, the existence of the predicted bi-stability jump regarding \mdot
  is not yet conclusive and an independent investigation of it is needed.
 
  Recent massive star evolution models, such as the Geneva models of \citet{groh14}, 
  show that most of the mass loss occurs due to stellar winds in the later supergiant phase, rather than during the early O-star phase near the zero-age main sequence (ZAMS). 
  
  The reason that O star mass-loss rates have been reduced is that empirical O star mass-loss rates have been down-revised by a factor of 2-3 
  due to wind clumping \citep{repolust04,mokiem07,shenar15}. 
{\bvp Currently, there is an ongoing debate whether the theoretical 
O star mass-loss rates are correct. The outcome will strongly depend on future progress in modelling optically thin and thick clumping in radiative transfer and alternative wind-modelling, as performed in the present paper. } 
}
The temperature dependence 
of the mass-loss rate is also of key importance when analysing the accuracy and suitability of mass-loss recipes for the effects of 
mass loss through stellar winds in massive star evolution. 

Whilst the existence of the \mdot jump is still under debate, other
data are unambiguous: the wind velocities of O stars and early
Bsgs are considerably larger those of the later Bsgs. 
This is likely to imply significant differences in
mass-loss rates of Bsgs in both temperature domains. It is matter of adequate
interpretation of the observations and/or treatment of the models to establish whether the trends are gradual \citep{cro06} or steep \citep{lamers95,vink99}.

\citet{lamers95} and \citet{vink99} suggested that there might be also 
  be a second bi-stability jump near 10\,000\,K which has not been studied in detail
  yet. In this paper, we investigate whether the second bi-stability jump exists, in conjunction 
  with studying the behaviour of the first bi-stability jump near
  $\sim$21\,000\,K.  Whilst the second jump could provide new insights into
  the properties of late B and A supergiants, for LBVs both bi-stability jumps are relevant, as their \Teff changes over a range
  between $\sim 8\,000$ and $30\,000$\,K
  \citep[\eg][]{gen01,vink12}. Therefore, if we understand the temperature
  behaviour of \mdot, we might be able to explain some of the mass loss variations
  during the different LBV phases \citep{vink02}. However, the
  effects of luminosity, clumping and metallicity on \mdot needs to be
  properly understood.

 The contents of the paper is as follows. In
    \S~\ref{wind_modelling} we briefly describe the method that we use
    to predict \mdot as well as our grid of models. The main results of
    current investigation are outlined and discussed in
    \S~\ref{sec:results}, whilst in \S~\ref{sec:comp} we compare mass loss rates following from the  \cmfgen models to the mass-loss rates resulting from the Monte Carlo calculations of \citet{vink00,vink01} (V00/V01).
    In \S~\ref{sec:LBVs}, we discuss the
    importance of the bi-stability jumps in \mdot for the behaviour of
    LBVs. Finally, in \S~\ref{sec:conclusions} conclusions are drawn.

%%%%%%%%%%%%%%%%%%%%%%%%%%%%%%%%%%%%%%%%%%%%%%%%
%METHODS
%%%%%%%%%%%%%%%%%%%%%%%%%%%%%%%%%%%%%%%%%%%%%%%%
\section{Wind modelling}
  \label{wind_modelling}

  Our current investigation utilises sophisticated fully line-blanketed 
  atmosphere models computed with the 
  non-LTE radiative transfer code \cmfgen
  \citep{hillier98}. \cmfgen is
  designed to solve the statistical equilibrium and radiative transfer
  equations in spherical geometry. Thus, the code is able to calculate
  the mass absorption coefficient $k_\nu$, and the total line acceleration,
  $g_{\rm L}^{\rm tot}$, which can be integrated directly:
    \begin{equation}
    \label{eq:line_acc}
     g_{\rm L}^{\rm tot}=\frac{1}{c}\int_{0}^{\infty}F_\nu k^{\rm L}_\nu \d\nu,
    \end{equation}
  where the stellar flux $F_\nu$ and $k^{\rm L}_\nu$ are computed on a relevant frequency grid. 

    When \gltot is known one may predict mass-loss rates of massive stars, as their winds are driven by the radiative pressure in spectral lines.
    
     \begin{table*}
	\caption{Prescribed stellar and wind parameters for the grid of models.}  
	\begin{center}	  
	  \begin{tabular}{ c c c c c c c  } 
	    \multicolumn{6}{c}{} 
	    \\ \hline\noalign{\smallskip}
	    log\,L/L$_\odot$ &${\rm M}_\star $    & $\Gamma_e$ & $Z/Z_\odot$ &\Teff  & \mdot  & model \\
	    & $({\rm M}_\odot)$  &            & & (K)            & ($10^{-6}\,{\rm M}_\odot\,{\rm yr^{-1}}$) & series\\
	    \hline\noalign{\smallskip} 
%	    5.00 & $13, 9.5, 6.3$ & $0.19, 0.26, 0.39$ & 0.2, 0.5, 1 & $8\,800-30\,000$ & 0.1 -- 0.75 & L5.0\\ 
            5.50 & $40, 30, 20 $  & $0.19, 0.26, 0.39$ & 0.1, 0.2, 0.5, 0.75, 1 & $8\,800-30\,000$ & 0.1 -- 2 & L5.5\\
            5.75 & $71      $	  & $0.19$             & 0.2, 0.5, 1 & $8\,800-30\,000$            & 0.1 -- 2 & L5.75M71\\ 
	  \hline    
	  \end{tabular}
	  \label{tab:grid} % is used to refer this table in the text
        \end{center}      
      \end{table*}

\subsection{\Q\,-- a tool to determine \mdot}      
      
  In order to predict \mdot, we compare the radiative energy lost due
  to all line-interactions, which is used to lift the mass out of
  the potential well and to accelerate the wind to \vinf, $W_{\rm
    wind}$:
  \begin{equation}
  \label{eq:2}
   W_{\rm wind}=\mdot \int_{R_\star}^{\infty}\left(g_{\rm rad}-\frac{1}{\rho}\frac{{\rm d}p}{{\rm d}r}\right){\rm d}r,
  \end{equation}  
  to the total energy of the wind, $L_{\rm wind}$ \citep[see for details][]{abbott85,vink99}:
  \begin{equation}
    L_{\rm wind}=\mdot\left(\frac{1}{2}\vinf^2 + \frac{G{\rm M}_\star}{R_\star} \right).
  \end{equation}
  This is similar to the work ratio method, $\Q=W_{\rm wind}/L_{\rm
  wind}$, used by \citet*{grafener02,grafener05} to check the
  consistency of their models. {\bvpb Note that in Eq.~\ref{eq:2}, $g_{\rm rad}$ does not only refer to line-interactions, but it includes all processes which contribute to the radiative
  acceleration (\eg Thomson scattering, bound-free acceleration)}.

  If the radiative acceleration provides sufficient force to lift a
  certain amount of mass, \mdot, out of the potential well, then $\Q$
  is expected to be of order unity. If $\Q<<1$, then the radiative acceleration cannot
  support a wind with mass-loss rate, \mdot.
  
  It should be stated that \cmfgen does not currently solve the wind momentum
  equation. Therefore, the wind velocity structure has to be
  prescribed. In our models, the wind velocity structure is described
  by a standard $\beta$-type velocity law with $\beta=1$, which is
  joined to the hydrostatic part of the wind just below the sonic
  point. The prescribed velocity structure might differ from the one
  obtained from the radiative acceleration. However, for different
  velocity laws one might obtain better agreement between the adopted and
  acquired velocity structures. Therefore, \Q still provides a meaningful
  criterion to establish whether the stellar wind can be driven, and
  used as an approximate tool to determine \mdot from models with
  specific stellar parameters.
  
  %For a given set of stellar parameters we predict the expected \mdot by adopting several values for the input \mdot and calculating model atmosphere with \cmfgen. For each set of stellar parameters we check for which of the adopted mass-loss rate is gives $\Q$ is expected to be 1. 

\subsection{The grid of models}

    \begin{figure}
	\centering        
	\resizebox{0.8\hsize}{!}{\includegraphics{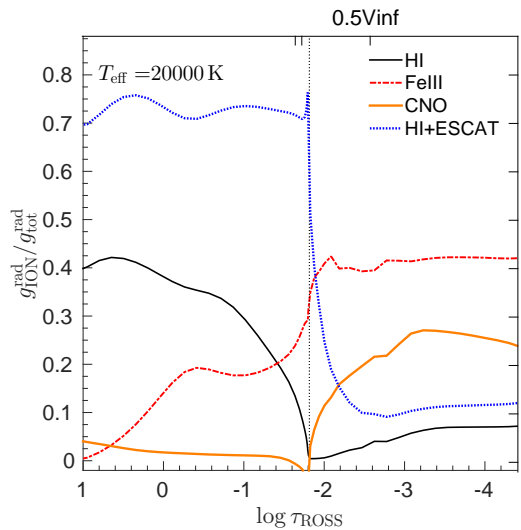}}
	\caption{Relative contribution of individual ions to the total radiative force for model from model series 'L5.5M40' with $\mdot=0.5\times10^{-6}\,\msunpyr$. The largest part of the \HI acceleration is due to bound-free processes}. 
	\label{fig:ACCvsTAU}
    \end{figure}

  We have calculated a grid of wind models covering a range of \Teff
  and \logg appropriate for blue supergiants, which include the ions of C,
  N, O, Ne, Mg, Al, Si, P, S, Ar, Ca, and Fe (with model atoms
  summarised in Table~\ref{tab:atm_data}). Observations indicate
    that stellar winds are inhomogeneous, and therefore modelling
    non-homogeneous stellar atmospheres becomes necessary.
  
  Currently \cmfgen is only able to take optically thin (micro) clumping into
  account. This micro-clumping approach is based on the hypothesis that the
  wind consists of small-scale over-density ``clumps'' which are
  optically thin.  The density $\rho$ within these clumps is assumed
  to be enhanced by a clumping factor $D$ compared to the mean wind
  density $\bar{\rho}$. This factor can also be understood in terms of
  volume filling factor $f=D^{-1}$, assuming that the inter-clump
  medium is void. Most of our models are non-homogeneous and are
  calculated with $f_\infty=0.1$, described by the following
  exponential law:
  \begin{equation}
    f(r) = f_{\infty} + (1 - f_{\infty}) \rm {exp}(-\varv(r)/\varv_{cl}),
  \end{equation}
  where $\varv_{\rm cl}$ is the velocity at which clumping is switched on. 
  We have chosen the clumping to start at $\varv_{\rm cl}=30$ km\,s$^{-1}$, i.e. just above the sonic point.
  In \S~\ref{sec:BSJ} however, we calculated a set of
    homogeneous models in order to estimate the influence of micro
    clumping on the bi-stability jump.
  
  To investigate whether the dependence of \mdot on \Teff is universal, 
  we calculated model series with different luminosities and masses in
  such a way that the {\it classical} Eddington parameter, $\Gamma_e$,
  is unchanged. 
  Unless otherwise stated, our \cmfgen models have been computed
  for 1/2 solar metal abundances\footnote{The reference solar abundances of
  \citet{asplund09} were adopted.}, in order to be relevant for results of the
   VLT-FLAMES Tarantula Survey \citep{evans11,bestenlehner14,mcevoy15}.
%  Following \citet{vink01}, 
We adopt the parameters summarised in Table~\ref{tab:grid}. 
For the main set of models we select masses  ${\rm M}_\star=40, 30$, and $20\,{\rm M}_\odot$ in order 
to compare our results to the \mdot calculations of \citet{vink01} and
to have the rough possibility to have Bsgs as 
core hydrogen burning main-sequence stars  as
well as core helium burning stars \citep[see \eg][]{vink10}. For the models with log\,$\Lstar/\Lsun=5.75$ we select ${\rm M}_\star=71\,{\rm M}_\odot$ 
in order to achieve the same mass to luminosity ratio as the models with log\,$\Lstar/\Lsun=5.50$  and ${\rm M}_\star=40\,{\rm M}_\odot$.  

  The wind terminal velocities, \vinf, were adopted according to the
  observed relations between \vinf and \vesc, \ie $\vinf/\vesc=2.6$
  for $\Teff\gtrsim21\,000$\,K, $\vinf/\vesc=0.7$ for
  $\Teff\lesssim10\,000$\,K and $\vinf/\vesc=1.3$ for \Teff in between
  \citep{lamers95,cro06,markova08,garcia14}.  Here, the escape velocity is
  given by:
  \begin{equation}
    \vesc=\sqrt{\frac{2G\Mstar}{{\rm R}_\star}},
  \end{equation}
  where \Mstar is mass of the star, and R$_\star$ is its radius.
  
 The definition of \vesc here is not corrected for the radiation pressure by electron scattering and therefore is different from the one used by
 \citet{vink99}.
 %Most investigations of the winds from massive stars work with effective escape velocity, \ie \vesc include $\Gamma_e$. 
  Consequently, the applied terminal velocities in present work are somewhat overestimated, when the values of 2.6, 1.3 and 0.7 are used, and the predicted \mdot might be underestimated. Overall, however, the influence should be small, except for those models with large $\Gamma_e$.
 
 In the present investigation we did not intend to make quantitative predictions. 
 {\bvpb Our models do not reach a similar level of completeness as those calculated by \cite{vink99}. 
 Whilst we might be missing the high-lying Fe lines and other elements from the Periodic Table in the current investigation, \cite{vink99} pre-selected the important driving lines out of a total line list of millions bound-bound transitions from \cite{kurucz88} from the first 30 elements in the Periodic Table (H - Zn). In other words, with the Monte Carlo method \cite{vink99} were {\bvpc fairly complete} and until we have similar level of completeness, we will refrain from making quantitative radiative acceleration predictions.
}
Instead, the aim of current paper is to investigate {\it qualitatively} both bi-stability jumps in \mdot that were predicted by \citet{vink99}.

%%%%%%%%%%%%%%%%%%%%%%%%%%%%%%%%%%%%%%%%%%%%%%%%
%RESULTS
%%%%%%%%%%%%%%%%%%%%%%%%%%%%%%%%%%%%%%%%%%%%%%%%
\section{Results}
  \label{sec:results}

  \subsection{Identity of the main line drivers}

    \begin{figure*}
	\centering        
	\resizebox{0.73\hsize}{!}{\includegraphics{Qtot_L5.5M40_0.5}\hspace{1.8cm}\includegraphics{Qrat_L5.5M40_0.5}}
	\caption{{\it Left}: \Q vs \Teff for models from 'L5.5M40' series with fixed mass-loss rate of $\mdot=0.5\times10^{-6}\,\msunpyr$. Observed velocity ratios of $\vinf/\vesc=2.6$ for $\Teff\ge22\,500$\,K, $\vinf/\vesc=1.3$ for $\Teff\in [10\,000$\,K$,20\,000$\,K], and  $\vinf/\vesc=0.7$ for $\Teff<10\,000$\,K are applied. {\it Right}: relative contribution of individual ions to the corresponding work ratio, \Q.}
	\label{fig:Q_L5.5M40_0.5}
    \end{figure*} 

    By calculating \Eq\ref{eq:line_acc} with the opacities of 
    different ions, we were able to extract the contribution of individual ions to the total %{\bvpc radiative} 
    acceleration.
    As an example, in Fig.\,\ref{fig:ACCvsTAU}, we
    show the relative contribution of individual ions to the total
    radiative {\bvpc force} for a model from the 'L5.5' series with mass, ${\rm M}_\star=40\,{\rm M}_\odot$ ('L5.5M40'). The model has $\Teff=20\,000$\,K,
    $\mdot = 0.5\times10^{-6}\msunpyr$.

    Close to the star, most of the radiative force is provided by
    neutral hydrogen (\HI) and electron scattering (blue dashed
    line)\footnote{Radiative acceleration of \HI (black solid) is
      mainly determined by bound-free processes}, whilst the most
    important line driver in the outer wind is \FeIII. The presented
    contributions to the total radiative force are distance dependent
    and it is thus ambiguous to establish which of the ions provide
    most of the global wind acceleration. Therefore, in
    \Fig\ref{fig:Q_L5.5M40_0.5}, we investigate which ions contribute
    mostly to the work ratio, \Q.
    
    {\bvpb In our \cmfgen models, the location of the lower boundary where \Q is calculated is defined at $\tau_{\rm ROSS}=100$.
    However, if the lower boundary is set at wind velocities $\sim0.2\times\vinf$,  $W_{\rm wind}$ changes by only a few percent, \ie   $W_{\rm wind}$ is mainly determined in the super-sonic part of the wind.
    Also, the pressure contribution to $W_{\rm wind}$ (and thus to \Q) is only relevant in the sub-sonic wind-region. Consequently, the pressure term does not play an important role in the calculation of the total $W_{\rm wind}$ (\Q).}

    {\it The left hand-side} of \Fig\ref{fig:Q_L5.5M40_0.5} depicts the
    behaviour of \Q as a function of \Teff in model series
    'L5.5M40' with fixed \mdot. \Q is decreasing when \Teff is reduced between 30\,000
    and 25\,000\,K and also between 20\,000 and 10\,000\,K, whilst
    between 22\,500 and 20\,000\,K a discontinuity in \Q is
    produced. The reasons for this are two-fold: (i) a change in Fe
    ionisation balance; and (ii) the different velocity ratio,
    \vinf/\vesc, for the models on both sides of the B-supergiant
    domain. For low values of \vinf/\vesc ($\Teff\leq20\,000$\,K), $\Q=1$
    should be easier to achieve in comparison to models with higher
    values of \vinf/\vesc ($\Teff\geq22\,500$\,K).

    \subsubsection{Acceleration by CNO}

    {\it The right-hand side} of \Fig\ref{fig:Q_L5.5M40_0.5} displays the
    relative contribution of individual ions to \Q. For simplicity,
    only the contributions of important ions are presented.  On the
    hot side of the bi-stability jump (at $\Teff\gtrsim22\,500$) the
    ions of C,N, and O contribute mostly to \Q. Thus, they contribute
    mostly to the line acceleration and they are important line drivers in
    hotter models. However, when \Teff is reduced from 25\,000 to
    20\,000\,K, the contribution of CNO to \Q decreases in a favour of
    Fe group elements (chiefly \FeIII).  %This is in agreement with previous results \citep[\eg][]{abbott82,vink99}. 
    {\bvp A similar finding was reported by \citet{abbott82}, but he did not make a distinction between the inner and outer wind.
    \citet{vink99} used the so-called sonic point, the  
    point at which the wind reaches the local speed of sound, 
    to distinguish the inner from the outer wind. 
    {\bvpb \cite{vink99} used this point as a physical point beyond which the mass-loss rate is already fixed but the wind terminal 
    velocity has yet to be determined \citep[see also][]{puls00}.
    }}

    \subsubsection{Iron - the wind driver}

    \begin{figure*}
	\centering        
	\resizebox{0.80\hsize}{!}{\includegraphics{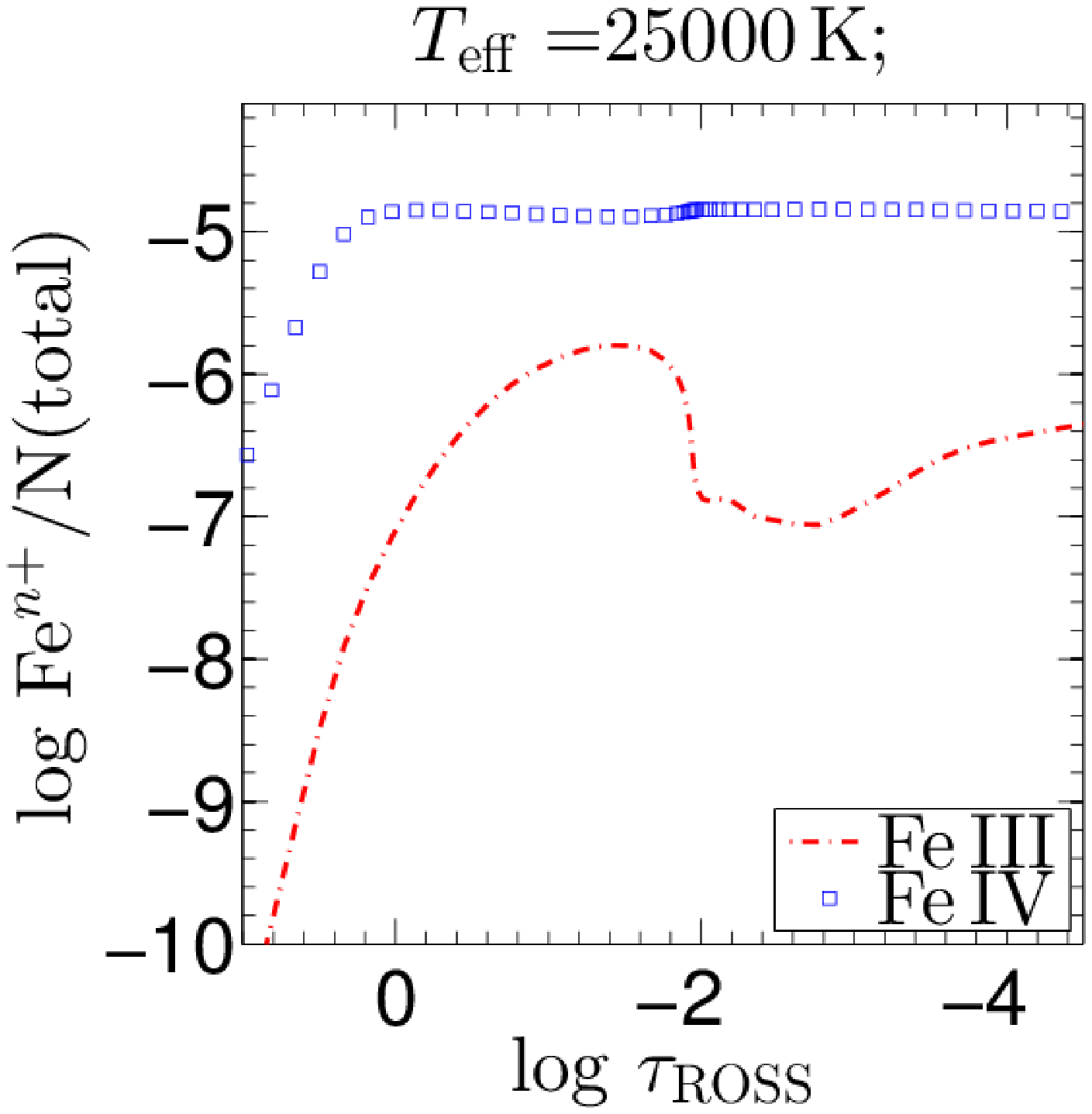}\hspace{0.45cm}\includegraphics{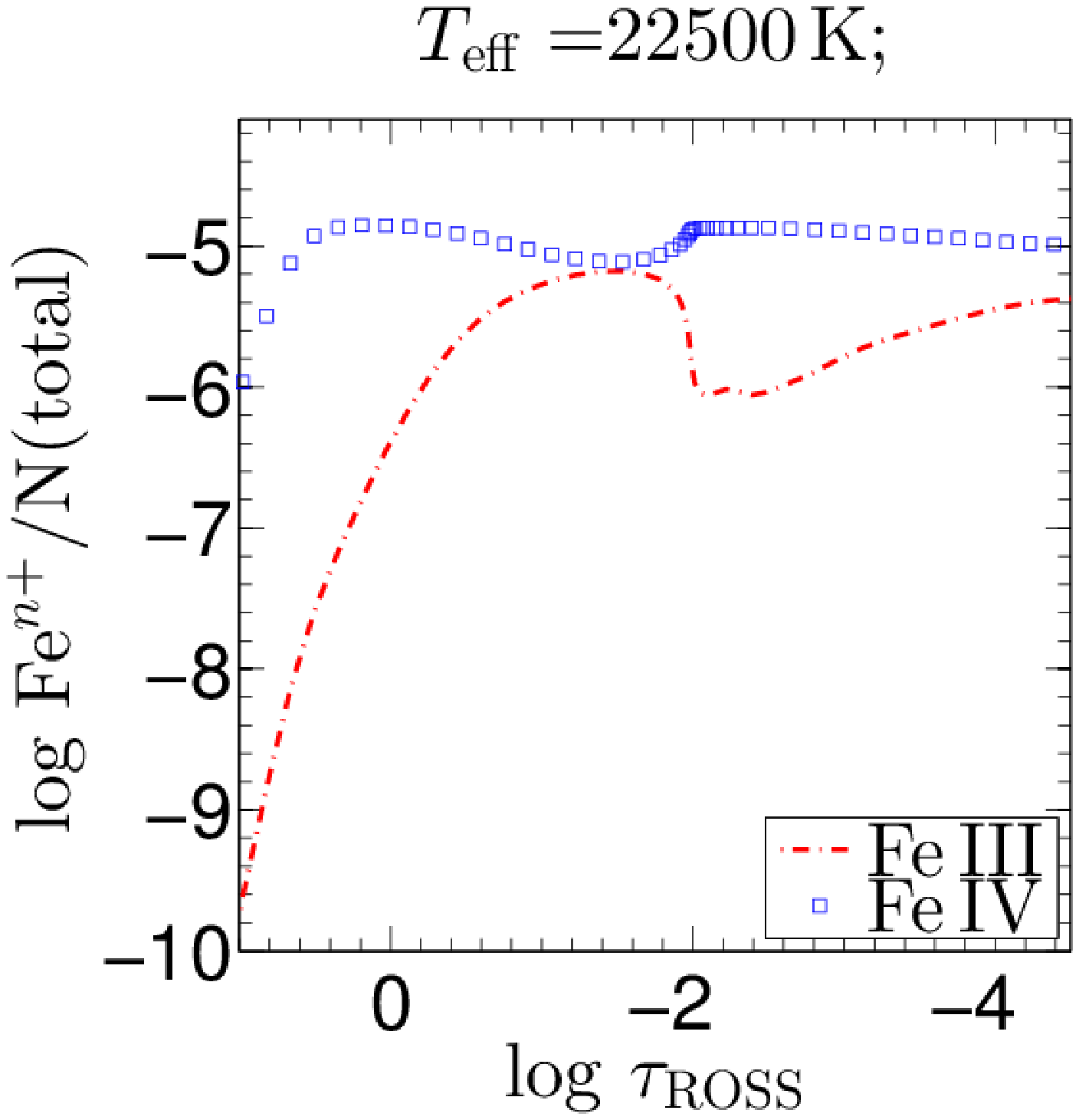}\hspace{1.0cm}\includegraphics{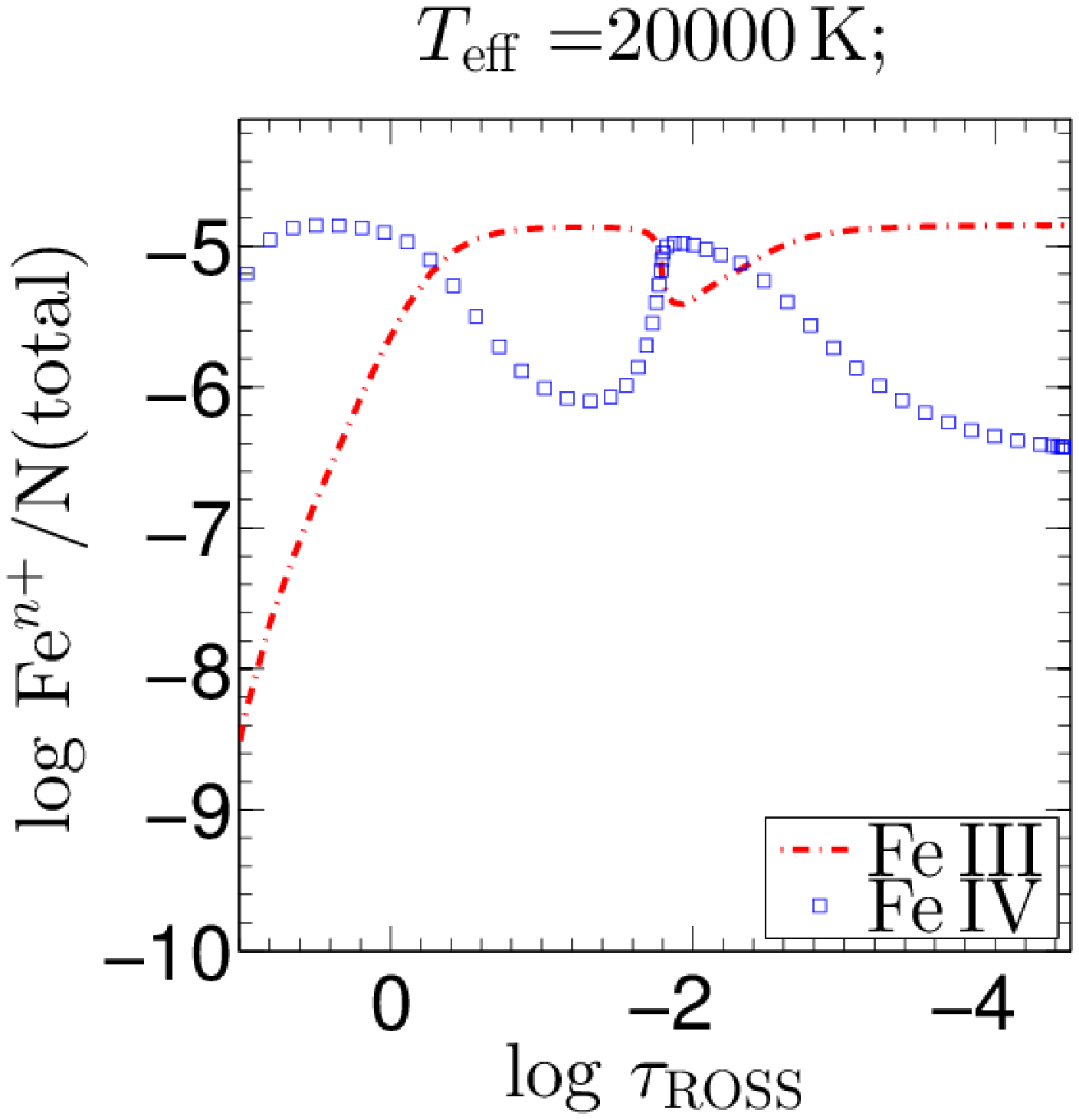}}
	\caption{Change in ionisation balance between \FeIV and \FeIII in models from 'L5.5M40' series with $\mdot=0.5\times10^{-6}\,\msunpyr$.}
	\label{fig:if_IRON}
    \end{figure*}

    Figure\,\ref{fig:if_IRON} displays that when \Teff is reduced from 25\,000 to 22\,500\,K, \FeIV decreases in favour of \FeIII. Even though  \FeIV is still the dominant ionisation stage in the cooler model, most of the radiative force of iron comes from \FeIII ({\cf \it right-hand side} of Fig.~\ref{fig:Q_L5.5M40_0.5}). %Note that, as \Teff is reduced from 25\,000 to 22\,500\,K, the contribution of C, N and O to the total line acceleration (and thus to \Q)  decreases in favour of \FeIII. 
    When \Teff is further reduced to 20\,000\,K, \FeIII becomes the dominant ionisation stage and now provides about 40\% of \Q. 

    The reader should be aware that we have prescribed
    $\vinf/\vesc=1.3$ for models with \Teff between 20\,000 and
    10\,000\,K in line with the observations \citep[see \eg][]{markova08}. 
    Thus, the models on the
    cool edge of the bi-stability jump would achieve more ``easily''
    the prescribed wind velocities than those models at the hot side,
    where the velocities are higher. In addition, lower velocities
    would also lead to higher densities (continuity equation), which
    stimulate recombinations. Consequently, between 22\,500 and
    20\,000\,K, a jump in \Q is produced. Note that this jump has to
    be accompanied by a jump in \mdot as well, because the model with
    $\Teff=20\,000$\,K would be able to drive a stronger wind. This is
    in agreement with previous studies \citep{vink99,vink01}, although
    \cmfgen predicts a jump at $\Teff\simeq20\,000\,$K, whilst
    according to the Monte-Carlo calculations the jump is expected at
    somewhat hotter temperatures, at $\Teff\simeq25\,000$\,K.

    Monte-Carlo calculations now have an improved line driving
    treatment \citep[see \eg][]{muller08,muijres12} and therefore a
    discordance in temperature of 5\,000\,K between \cmfgen  and
    Monte-Carlo models is particularly intriguing. Such a large
    discordance may underline fundamental differences between the
    assumptions regarding the treatment of the ionisation in both
    codes, or the discordance might be caused by differences in the atomic
    data which both codes use. 
{\bvpb  The main difference between the \cmfgen models and the Monte-Carlo calculations is probably the 'exact' vs simplified Non-LTE treatment, which was
calculated by V00/V01 using the ISA-Wind code \citep{dekoter93}. This is the most likely reason for the discrepancies, as the ISA-Wind calculations do not include line blanketing and therefore the iron ionisation would occur at higher \Teff in comparison to \cmfgen models. }
    
  \subsection{Bi-stability jump on trial}
  \label{sec:BSJ}

    \cmfgen does not currently calculate mass-loss rates. Instead, \mdot is required as an input parameter. Nevertheless, as was discussed earlier, the \Q ratio enables us in some way to estimate for which value of \mdot, a specific model is able to drive its stellar wind. 
    For models with given set of stellar parameters we predict their mass-loss rate by adopting several values of \mdot.  The value of \mdot for which $\Q$ becomes unity is the predicted mass-loss rate.
    
\begin{figure*}
	\centering        
	\resizebox{0.9\hsize}{!}{\includegraphics{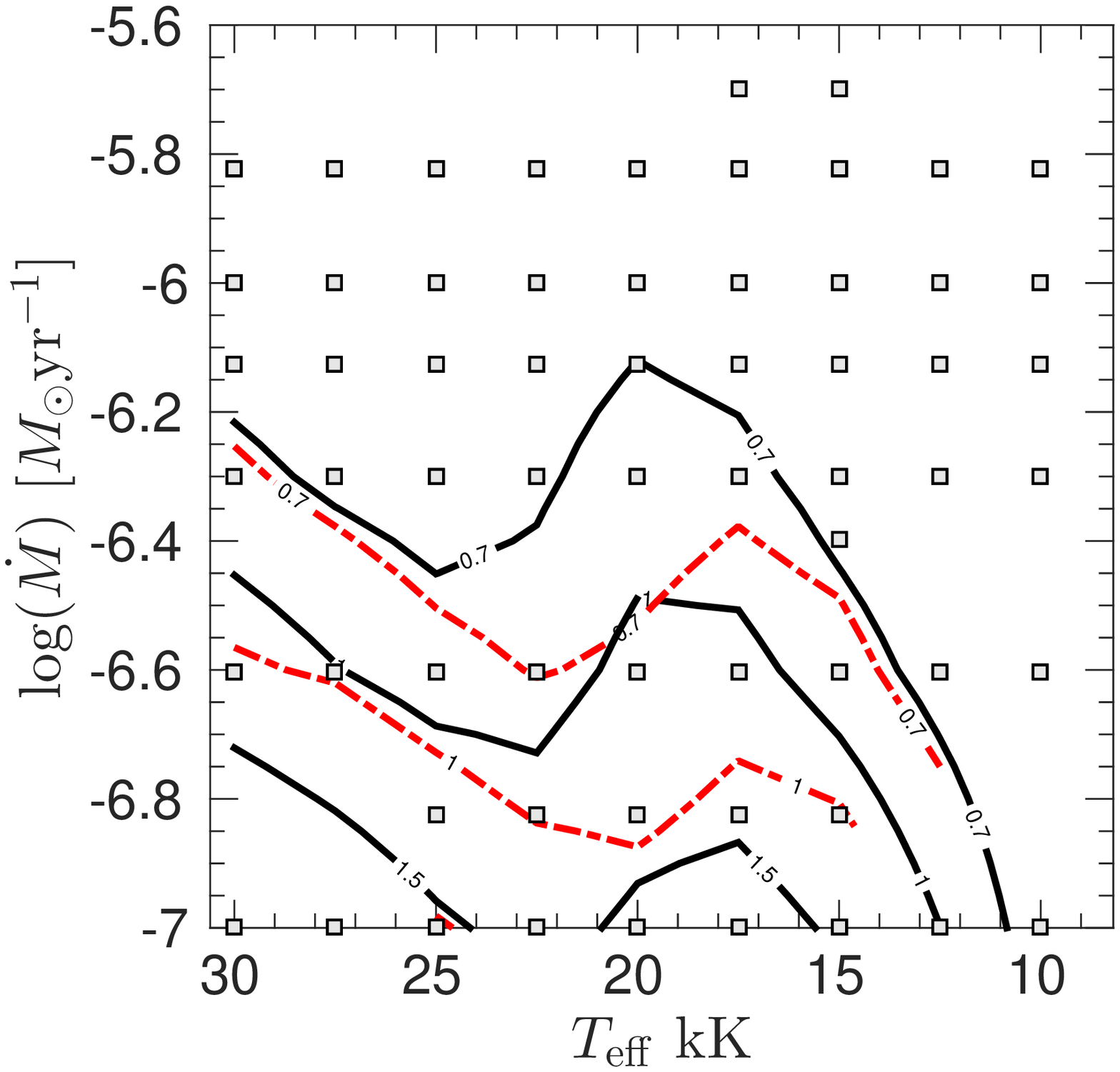}\hspace{-3.0cm}\includegraphics{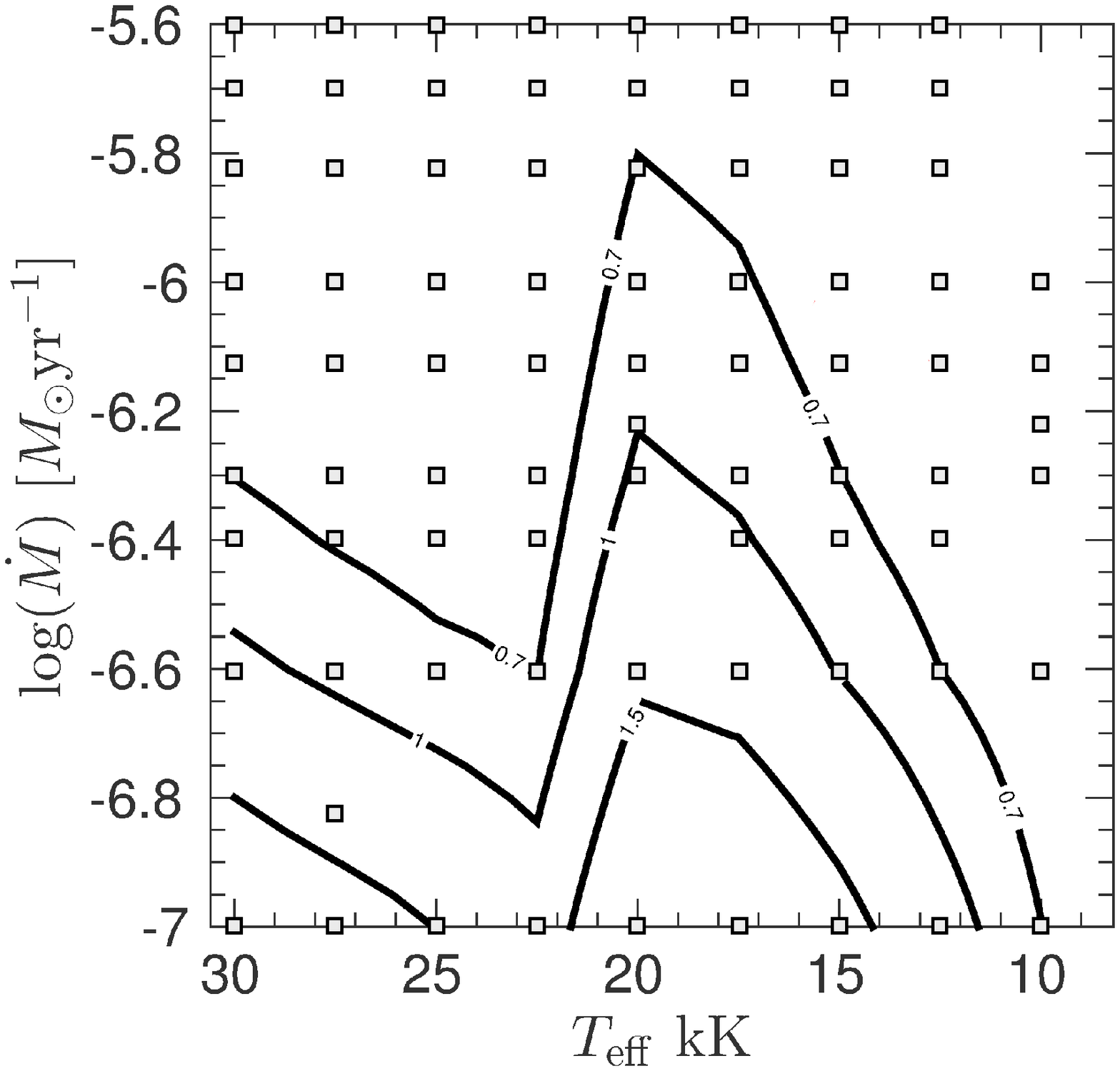}}
	\caption{{\it Left}: contour plot of \Q as a function of \Teff and \mdot in model series 'L5.5M40' with $\vinf/\vesc=2$. To see the effect of clumping, iso-contours from models with no micro-clumping taken into account are over-plotted with red dashed lines. %If $\Q\sim1$  then the radiative acceleration provides sufficient force to support a wind with mass-loss rate,\mdot.
	{\it Right}: contour plot of \Q from the same model series but with observed ratio of $\vinf/\vesc=2.6$ for $\Teff>\sim21\,000\,K$ and $\vinf/\vesc=1.3$ for $\Teff<\sim21\,000\,K$. White squares mark the positions of the of the calculated models.}
	\label{fig:BJ_L5.5M40}

	\centering        
	\resizebox{0.72\hsize}{!}{\hspace{-4cm}\includegraphics{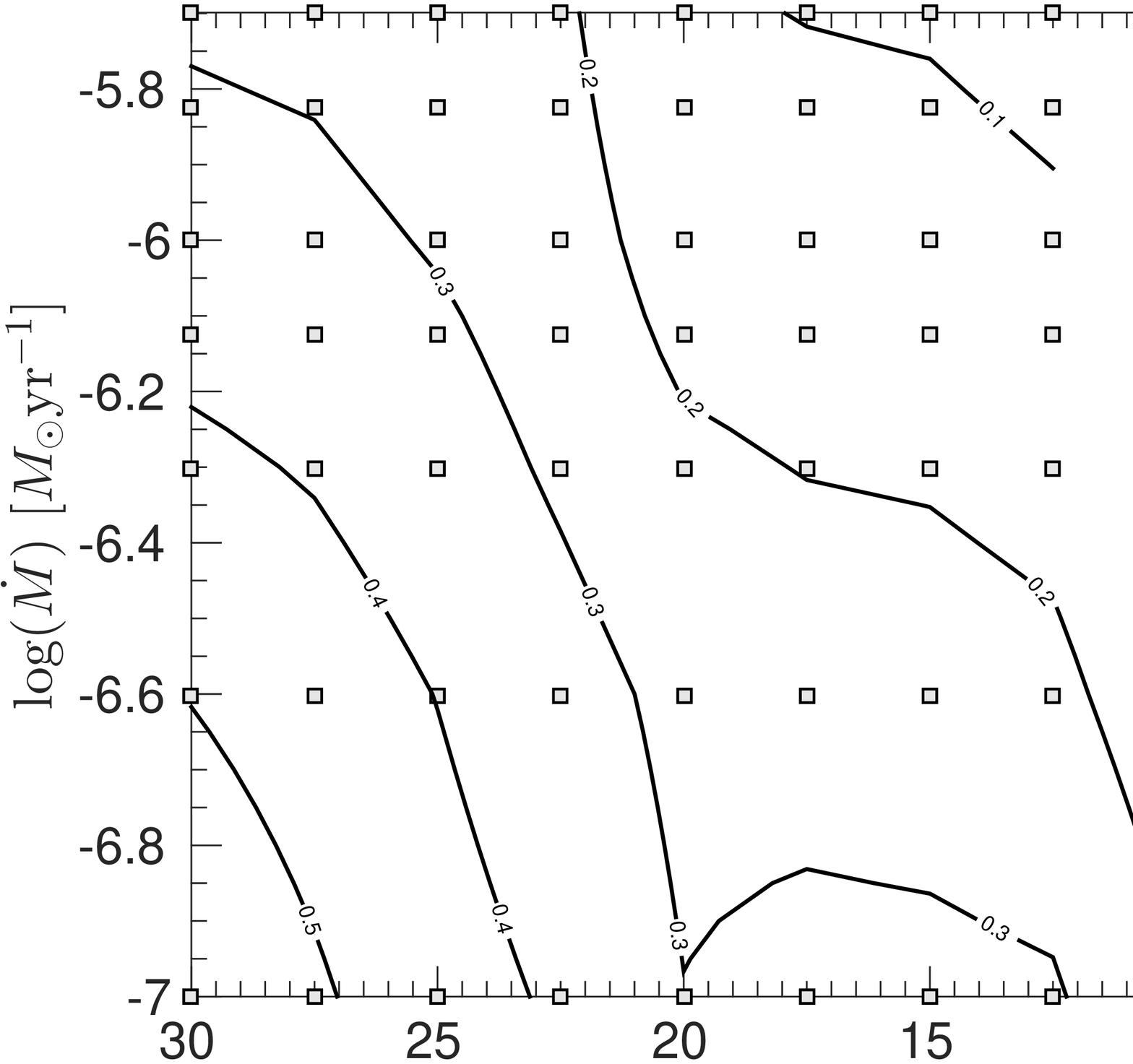}\hspace{2.0cm}\includegraphics{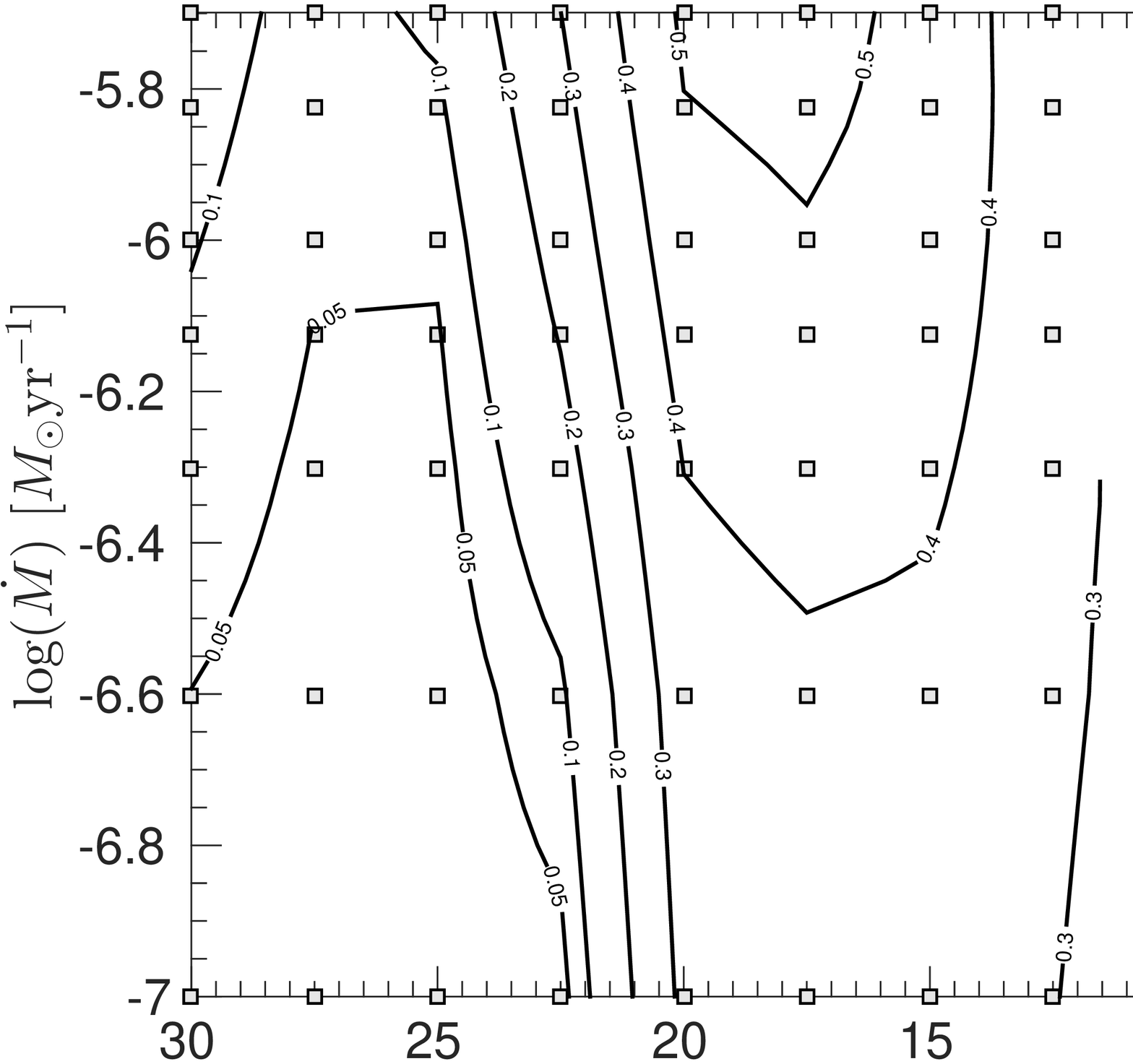}}
	\caption{{\it Left}: contour plot of the relative contribution of the ions of C,N, and O to \Q, $Q_{\rm CNO}$/\Q,  from model series 'L5.5M40'. {\it Right:} contour plot of the relative contribution of iron to \Q, $Q_{\rm Fe}$/\Q, for same models. For $\Teff\ge22\,500\,$K $\vinf/\vesc=2.6$, whilst for $\Teff\le20\,000\,$K $\vinf/\vesc=1.3$. }
	\label{fig:BJ_L5.5M40_CONTR}
    \end{figure*}

    To find out for which \mdot our models acquire  $\Q\approx1$, we investigate the behaviour of \Q as a function of \mdot and \Teff. On the {\it left hand-side} of Fig.~\ref{fig:BJ_L5.5M40}, we present a contour plot of \Q depending on \Teff and \mdot in model series 'L5.5M40' with $\vinf/\vesc=2$, $f_\infty=0.1$, and $\varv_{\rm cl}=30$\,km/s. The figure demonstrates that for a constant velocity ratio, between 22\,500 and 20\,000\,K, the predicted \mdot  is increased by a factor of about two. The temperature location of the predicted increase in \mdot is in good correspondence to observational findings \cite[see \eg][]{markova08}. If the observed velocity ratios are applied (\ie $\vinf/\vesc=2.6$ for $\Teff\ge22\,500$\,K and $\vinf/\vesc=1.3$ for \Teff between 20\,000 and 10\,000\,K), then \mdot is increased by about a factor of four (\cf\,{\it right-hand side} of Fig.~\ref{fig:BJ_L5.5M40}).
    On the basis of Fig.~\ref{fig:Q_L5.5M40_0.5} we confirm  that \FeIII is indeed responsible for this jump.

    In Fig.~\ref{fig:BJ_L5.5M40_CONTR}, we show contour plots of the
    relative contribution of the ions of C, N, and O ({\it left}), and
    iron ({\it right}) to the total \Q ratio. It is interesting to
    note that with an increase of \mdot the contribution of C, N, and
    O to \Q is decreased in favour of iron. When \mdot is increased
    by about three times in the models with
    $\Teff\sim20\,000-17\,500\,$K, the contribution of iron (chiefly
    \FeIII) to \Q is increased by about 25\% ($Q_{\rm Fe}$/\Q
    increases from 0.4 to 0.5). This is because in stronger winds the
    recombinations of \FeIV to \FeIII are favoured and the absolute
    number of iron ions increases.

    {\bvp The red dashed iso-contours on the {\it left-hand side of} Fig.~\ref{fig:BJ_L5.5M40} display the behaviour of \Q as a
    function of \Teff and \mdot in homogeneous sets of models with fixed velocity ratio of $\vinf/\vesc=2$. Note that micro-clumping has a
    large impact on \Q for temperatures between $\sim17\,500$ and $\sim22\,500$\,K, \ie around the bi-stability jump location. 
    For the other temperatures the influence of micro-clumping is less important.
    Moreover, the temperature at which bi-stability jump occurs in the homogeneous set
    of models ($\Teff=17\,500$\,K) is lower in comparison to the initial set of models with $f_\infty=0.1$, $\varv_{\rm cl}=30$\,km/s}. Thus, the degree of clumping seems to be an
    important parameter for the temperature location of the
    bi-stability-jump. We found that in model series with values for
    $\varv_{\rm cl}$ up to 200 km/s, the bi-stability jump occurs at
    $\Teff=20\,000$\,K and therefore $\varv_{\rm cl}$ has little
    effect on the temperature location of the bi-stability jump.  
  
  \subsection{A second bi-stability jump?}

    To investigate whether a second jump in \mdot exist near
    $\Teff=10\,000$\,K, we have calculated additional models with
    $\Teff=8\,800$ \,K\footnote{Below 8\,800\,K, a
      self-consistent hydrostatic solution in the hydrostatic part of
      wind was not obtained and therefore our grid stops at
      8\,800\,K.}. For $\Teff\lesssim10\,000$\,K \citet{lamers95} found
    that $\vinf/\vesc=0.7$ and therefore, we used such velocity ratio
    for those models. The reader should be aware that the terminal
    velocities of stars with temperatures below $10\,000$\,K were
    measured with an accuracy between 10\% and 20\% and therefore the
    value of $\vinf/\vesc=0.7$ might be uncertain.

    Nevertheless, we consider the adopted value as reasonable because:
    (i) in our coolest models the ions of \FeII provide most of the
    line acceleration (\cf\,{\it right hand-side of}
    Fig.~\ref{fig:Q_L5.5M40_0.5}), which is in agreement with previous
    investigations \citep[\eg][]{vink99,vink02}, and therefore \FeII
    could influence \vinf and \mdot; and (ii) the temperature range
    where \FeII becomes the main line-driver is between $\sim$10\,000
    and 8\,800\,K, which is the temperature range where
    \citet{lamers95} suspected that there might be a second
    bi-stability jump.

\begin{figure}
      \centering
	\resizebox{\hsize}{!}{\includegraphics{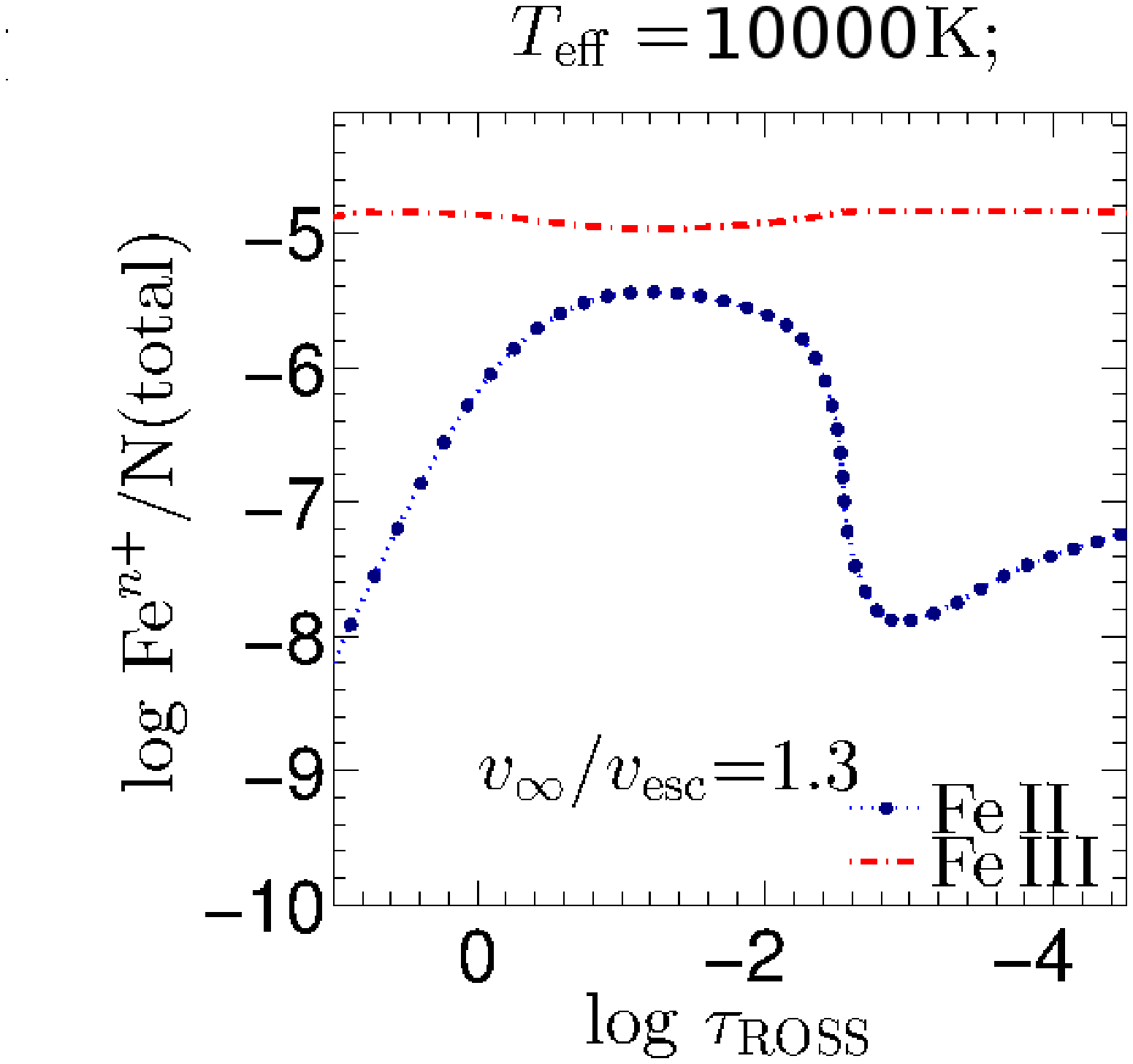}\hspace{0.0cm}\includegraphics{if_IRON_8.8e}}
	\caption{Change in ionisation balance between \FeIII and \FeII in models from 'L5.5M40' series with $\mdot=0.25\times10^{-6}\,\msunpyr$.}
	\label{fig:if_IRONII}
    \end{figure}

    Figure~\ref{fig:if_IRONII} reveals that when \Teff is reduced
    below 10\,000, \FeIII recombines to \FeII, similarly to the
    recombination/ionisation of \ion{Fe}{iv/iii} already presented in
    \Fig\ref{fig:if_IRON}. Note that at the coolest model, \FeIII is
    not fully recombined to \FeII. Whereas in the inner (sub-sonic) part of the
    wind \FeII is the dominant ionisation stage, in the outer wind
    \FeIII is still the dominant ion, even though \FeII contributes
    most to the total acceleration provided by iron, as shown in {\it
      right hand-side} of Fig.~\ref{fig:Q_L5.5M40_0.5}. If \Teff is
    further reduced, we anticipate \FeII to become the dominant ion
    throughout the wind and to provide an even larger fraction of the 
    radiative acceleration.

    The model with $\Teff=8\,800$\,K from 'L5.5M40' series does not achieve $\Q=1$ at all when \mdot is varied from $10^{-7}$ to $3\times10^{-6}$ \msunpyr. For such  \Teff, $\Q$ becomes of order unity when  metal abundances are increased (\cf \S~\ref{sec:met}) or  when models get close to the Eddington limit \citep{eddington1921}. 
    The proximity to the Eddington limit is characterised by the {\it classical} Eddington factor:
    \begin{equation}
     \Gamma_e=\frac{\sigma_e \Lstar}{4\pi c G {\rm M}_\star}
    \end{equation}
    {\bvpb where $\sigma_e$ is electron scattering opacity per unit mass and in the CGS system is measured in cm$^2$/g.}    
    In hot stars, $\sigma_e$ is constant throughout the wind, as the majority of 
    H and He ions are completely ionised. 
    {\bvpb Thus, for early and mid Bsgs the Eddington factor depends only on the ratio between stellar luminosity and mass. However, for late Bsgs and Asgs this is no longer true.
    }
    
    We constructed our models close to the Eddington limit by decreasing
    the stellar mass, whilst the luminosity was kept fixed. We selected masses
    ${\rm M}_\star=20$ and ${\rm M}_\star=30\,{\rm M}_\odot$ for model series
    'L5.5M20' and 'L5.5M30' respectively (\cf
    Table~\ref{tab:mdot_pred}).
    The models with $\Teff=10\,000$\,K obtain $\Q=1$ at
    $\mdot\approx0.09\times10^{-6}$\,\msunpyr and
    $\mdot\approx0.06\times10^{-6}$\,\msunpyr for
    ${\rm M}_\star=20$ and ${\rm M}_\star=30\,{\rm M}_\odot$ respectively. However, the coolest model from 'L5.5M30' series still does not achieve $\Q=1$ for \mdot in range between 0.2 and 2$\times10^{-6}$\,\msunpyr. The mass of the model has to be reduced to 20\,$\,{\rm M}_\odot$ in order for \Q to become unity. Thus, between 10\,000 and 8\,800\,K \cmfgen predicts a jump in mass-loss rate, \mj, only in models series 'L5.5M20'.

    According to Fig.~\ref{fig:Q_L5.5M40_0.5} for the model with
    $\Teff=8\,800$\,K \FeII contributes mostly to the work ratio: it
    provides nearly 70\% of \Q. Thus, the predicted second jump in
    \mdot should be caused by the radiative acceleration provided by
    \FeII. This implies that mass-loss rates of late B/A supergiants
    and LBVs are sensitive to the ionisation equilibrium of iron. The
    reader should be aware that \FeII is not fully recombined at
    $\Teff=8\,800$\,K, and therefore we expect at cooler temperatures
    \mj to increase {\bvpb further. This is very important for stellar evolution considerations.}
    
  \subsection{The influence of metal abundances}
  \label{sec:met}
  
      \begin{figure}
	\centering 
	\resizebox{\hsize}{!}  {\includegraphics{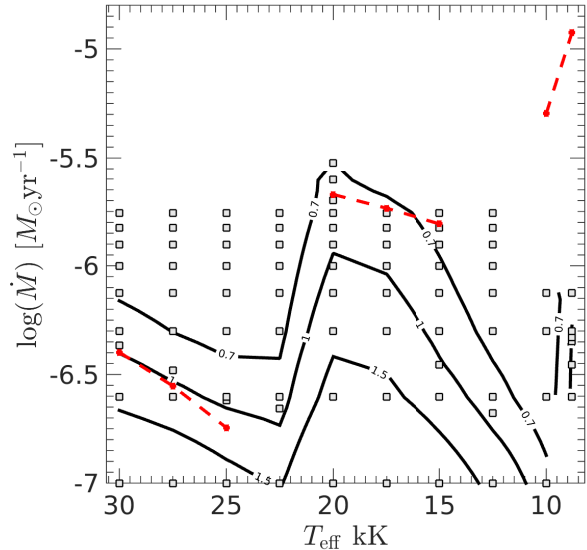}} % 0.75 old
	\caption{Contour plot of \Q in $\mdot-\Teff$ plane for model series 'L5.5M40'. Models have solar metal composition and parameters as listed in Table~\ref{tab:mdot_pred}. For comparison mass-loss rates calculated with the theoretical recipe of \citet{vink01} are shown with red dashed line. 
	}	
	\label{fig:GALBJ_L5.5M40_ISO}
    \end{figure}
    
  The magnitude of \mj depends on the metal composition. To investigate that, we have calculated a grid of models with solar metal abundances.
      Figure~\ref{fig:GALBJ_L5.5M40_ISO} compares contour plots of \Q in
  $\mdot-\Teff$ plane for model series 'L5.5M40' with
  solar metal abundances. The observed velocity ratios are applied.
  It is interestingly to note that in model series 'L5.5M40' between
  10\,000 and 8\,800\,K, \mdot at which $\Q=1$ increases from
  $\sim0.06$ to $\sim0.55\times10^{-6}\,\msunpyr$ (\cf
  Table~\ref{tab:mdot_pred}), whilst for half-solar metallicities the
  coolest models did not acquired $\Q=1$ at all. This implies that the
  second jump should be favoured in high metallicity environments, and
  for low metal environments, this jump is relevant only for
  objects close to the Eddington limit ($\Gamma\sim1$).
  
  {\bvpb
  If a constant velocity ratio of 1.3 is applied for the models with $\Teff=10\,000$\,K and $\Teff=8\,800\,$K {\bvpc and with half-solar metallicities}, a second bi-stability jump is
  produced only when the models are close to the Eddington limit (\ie 'L5.5M20' model series). In 'L5.5M30' \& 'L5.5M40' model series, a second BSJ is not produced at all when the constant velocity ratio of 1.3 is applied. This implies that the second BSJ in these series is a consequence of the applied observed velocity ratios.
  }
  
\begin{figure*}
        %{\begin{minipage}{\linewidth}
	%\centering
	  \vspace{-0.5cm}\hspace{-0.15cm}\includegraphics[width=0.49\linewidth, height = 0.9\textheight, keepaspectratio=true]{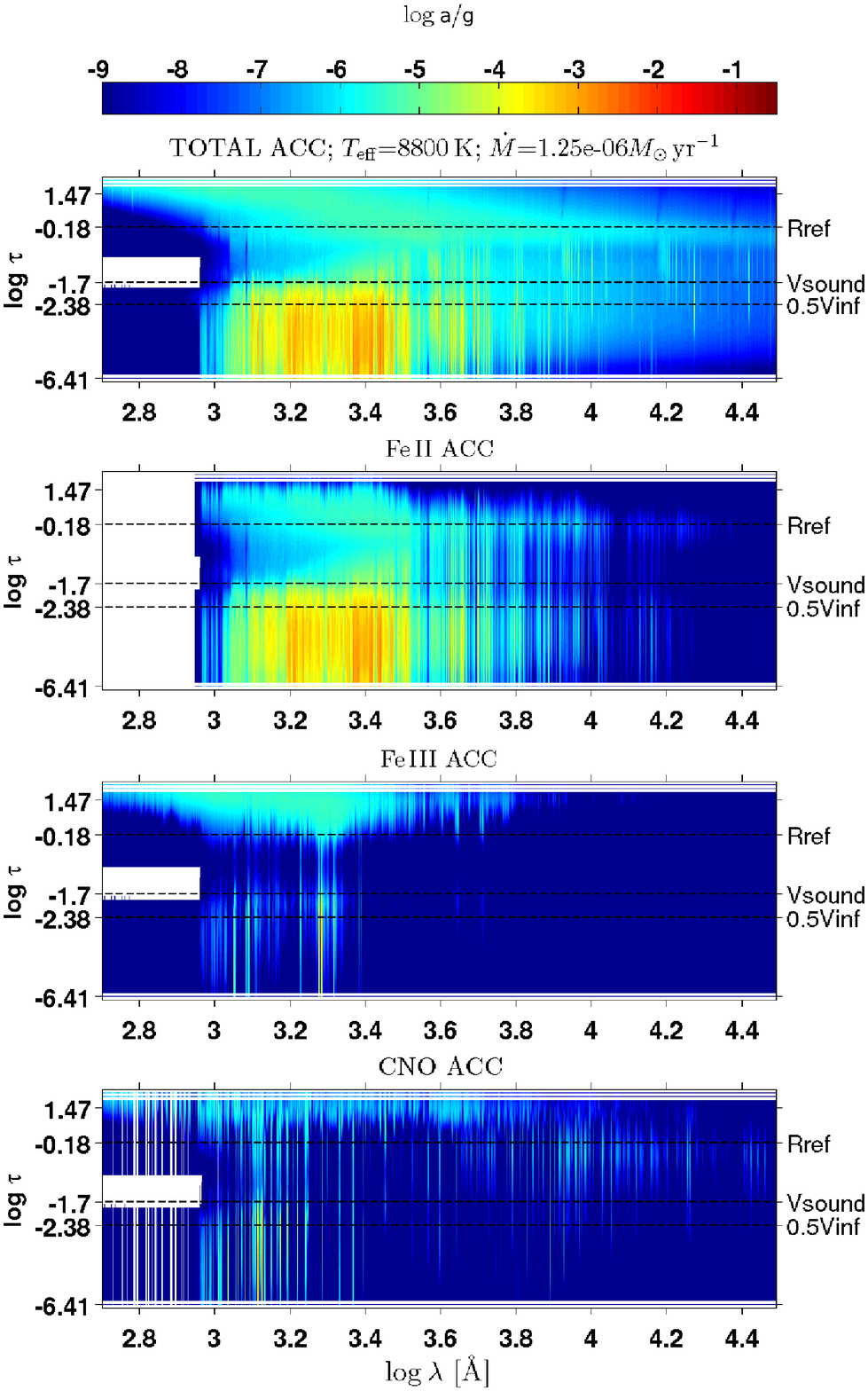}\hspace{0.2cm}\includegraphics[width=0.49\linewidth, height = \textheight, keepaspectratio=true]{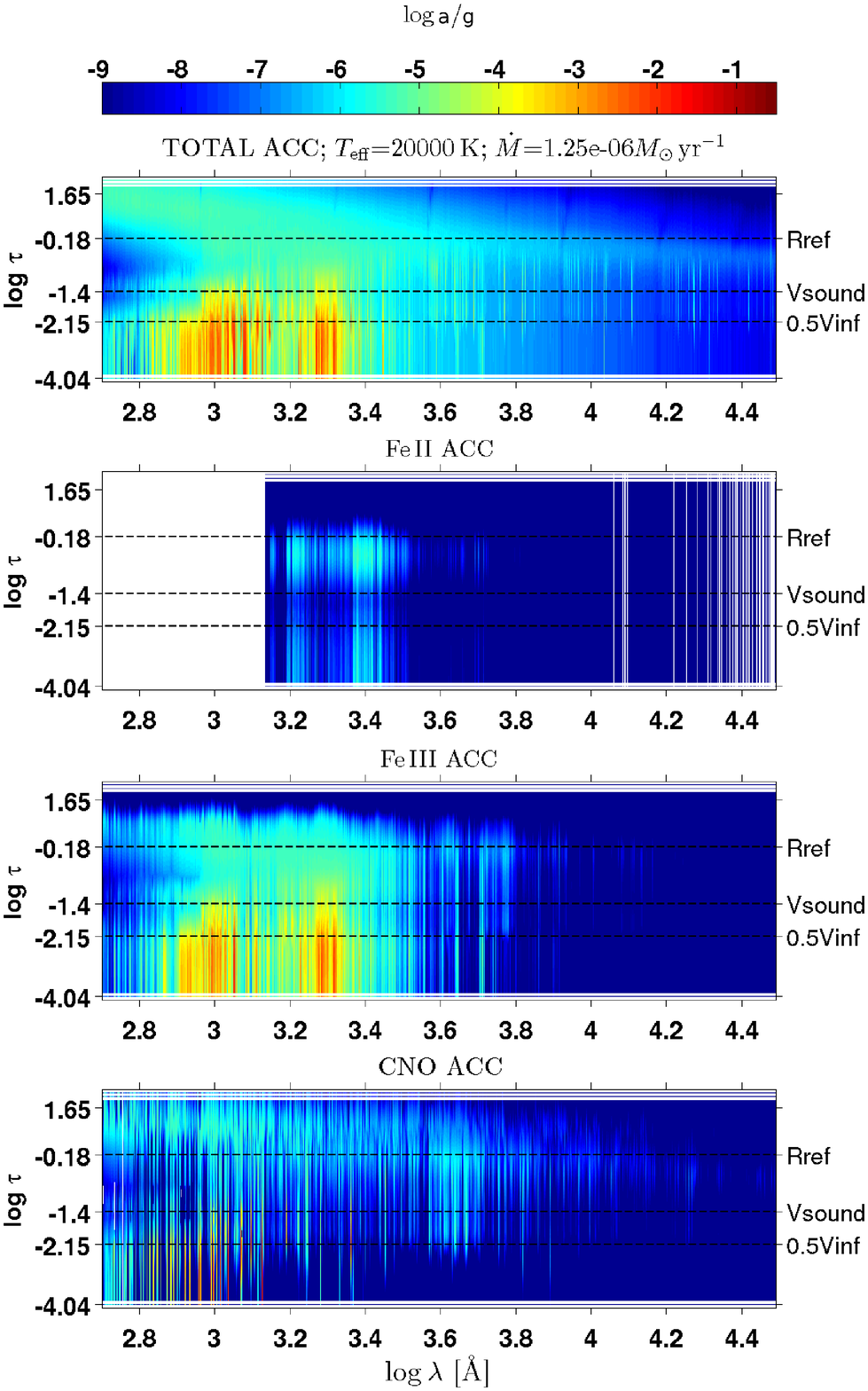} \vspace{0.16cm}
	                
                \vspace{-0.6cm}\hspace{0.1cm}\includegraphics[width=0.485\linewidth, height = \textheight, keepaspectratio=true]{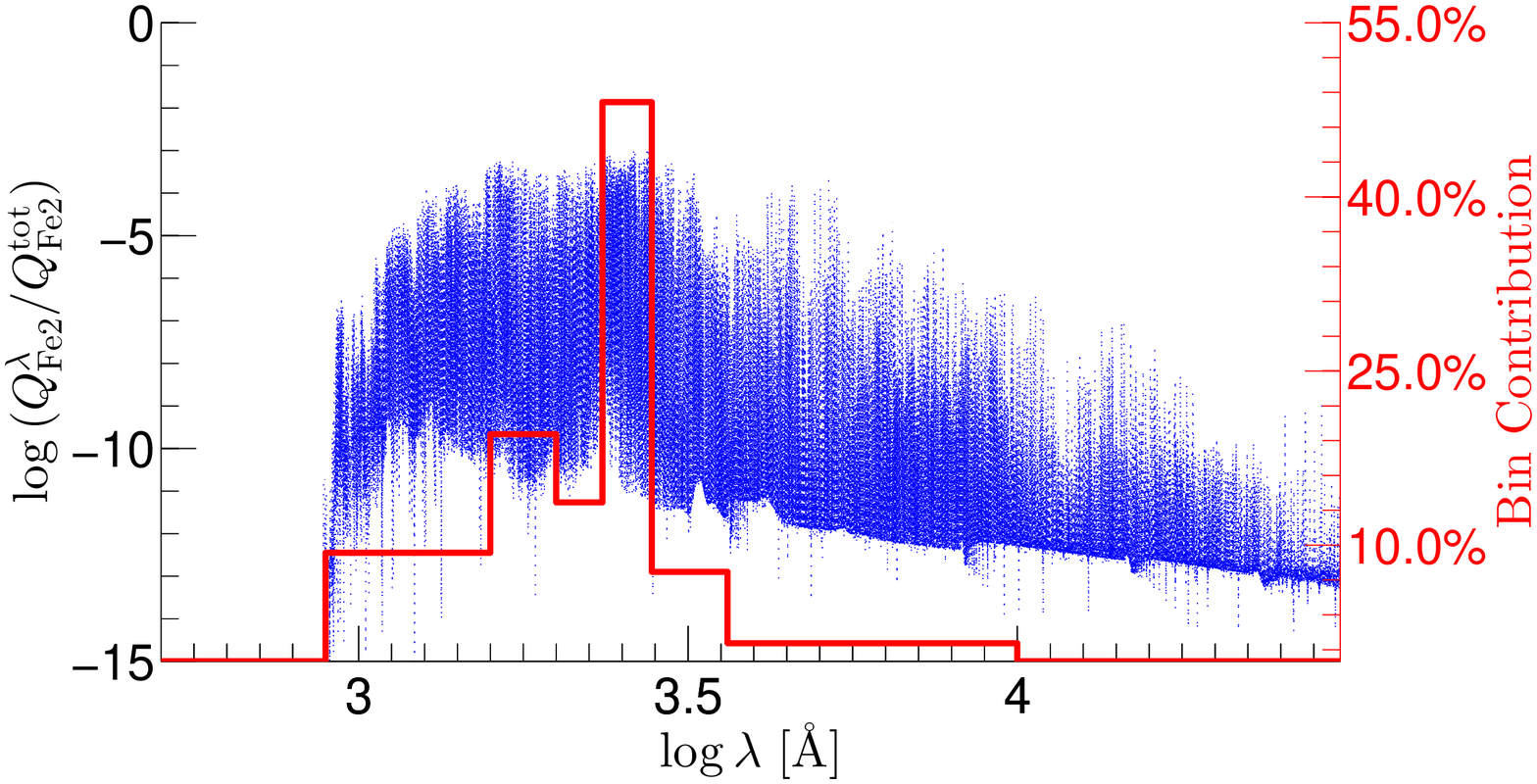}\hspace{0.38cm}\includegraphics[width=0.48\linewidth, height = \textheight, keepaspectratio=true]{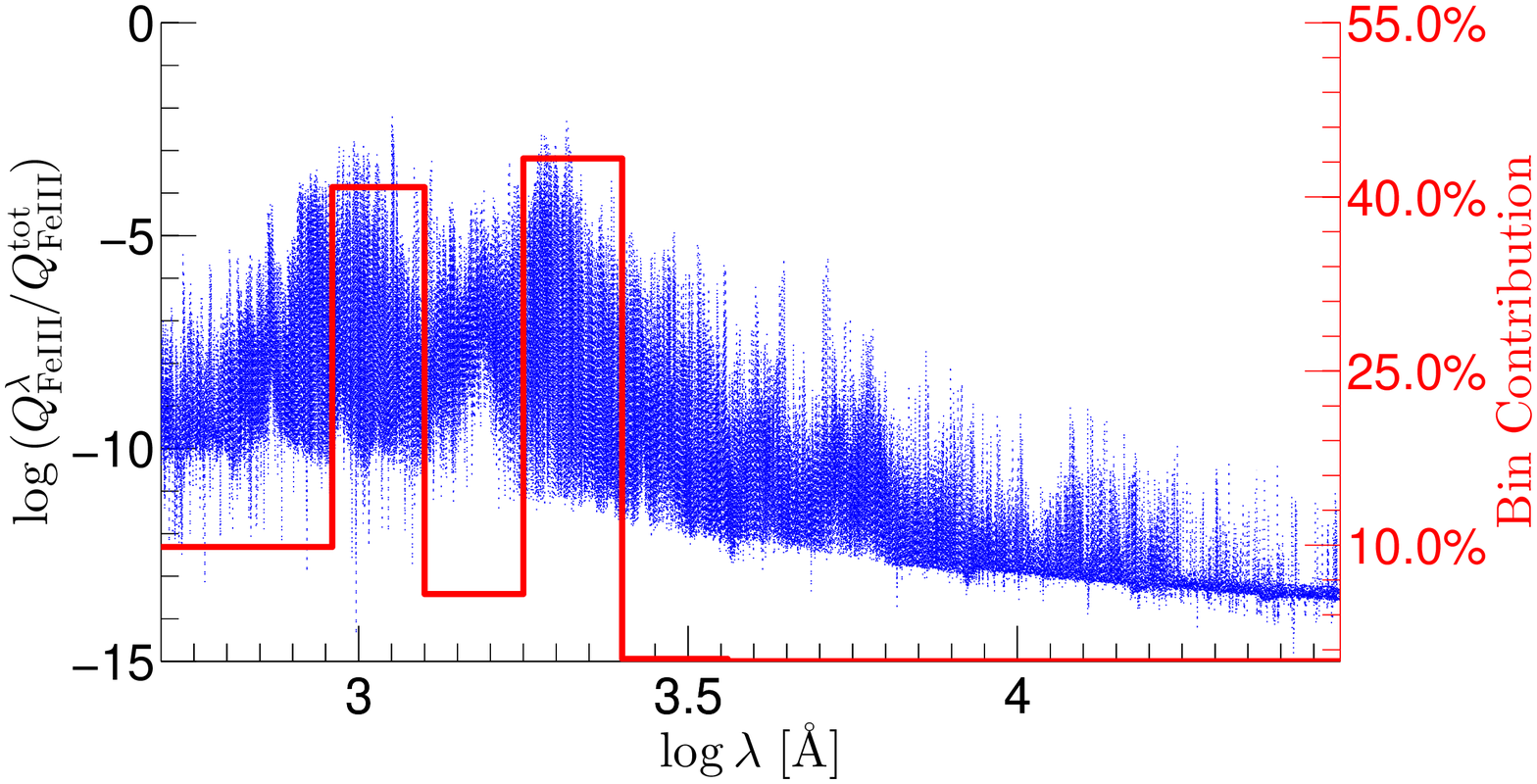} \vspace{0.16cm}

                \vspace{-0.2cm}\hspace{0.1cm}\includegraphics[width=0.48\linewidth, height = \textheight, keepaspectratio=true]{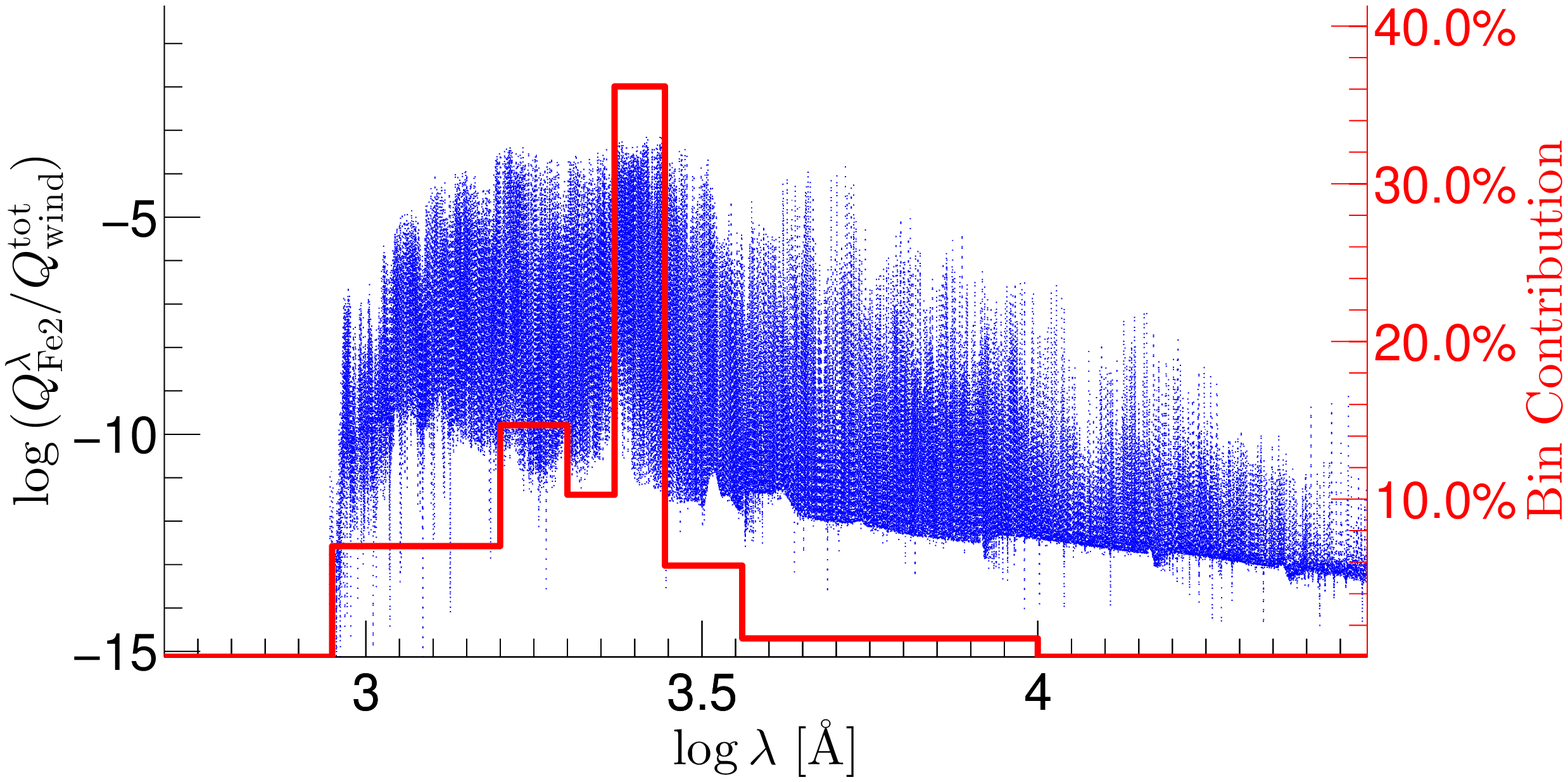}\hspace{0.38cm}\includegraphics[width=0.48\linewidth, height = \textheight, keepaspectratio=true]{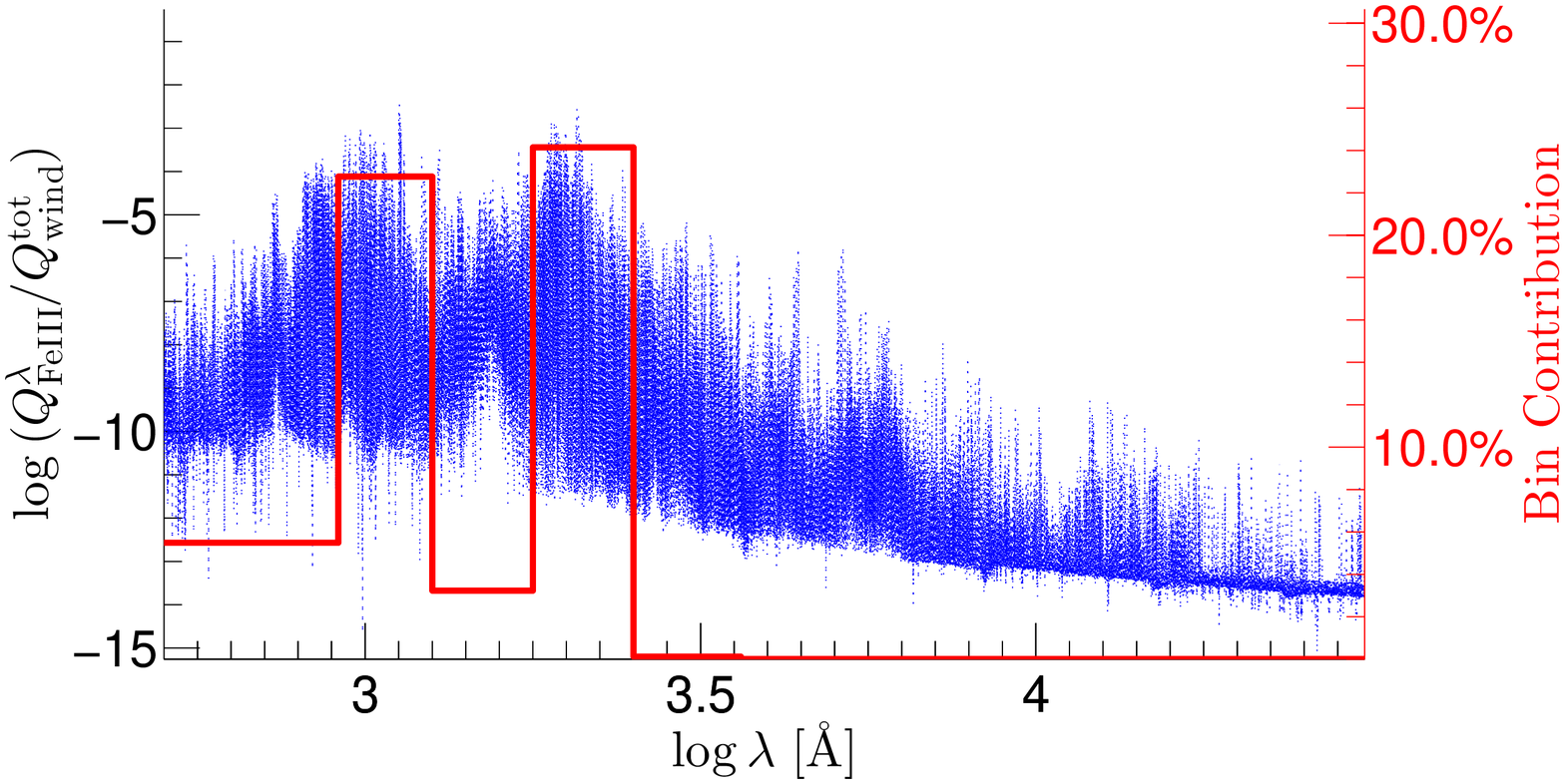}
         %\end{minipage}} 
	\caption{Radiative force provided by \FeII, \FeIII, CNO, and all ions as a function of $\lambda$ and $\tau_{\rm ROSS}$ for models with $\Teff=8\,800$\,K ({\it left}) and $\Teff=20\,000$\,K ({\it right}). Models have solar metal composition. The lowermost panels illustrate the contributions of the spectral lines to the work ratio obtained by the acceleration from \FeII (left) or \FeIII (right). The red line (with ordinate on the {\it right-hand side}) presents the total contribution of spectral lines located in various frequency bins to the work ratio of \FeII ({\it left}) or \FeIII ({\it right}).}
	\label{fig:ACC_3D}
    \end{figure*}

  To investigate in detail the origin of the second bi-stability jump,
  we show the total radiative acceleration in Fig.~\ref{fig:ACC_3D} in
  units of the local gravity (uppermost panels) as a function of $\lambda$
  and $\tau_{\rm ROSS}$ for models with $\Teff=8\,800$\,K ({\it left})
  and $\Teff=20\,000$\,K ({\it right}). For comparison, the radiative
  force due to the spectral lines of of \FeII, \FeIII, and CNO is
  displayed as well. In the cooler model, most of the radiative
  acceleration is provided by \FeII, whilst in the hotter model \FeIII
  is the dominant wind driver.
  
  Note that there are white regions present in Fig.~\ref{fig:ACC_3D} 
    where no information about radiative acceleration is
    revealed. Whilst in the second panel of the figure, the white
    regions are due to the lack of line transitions of the \FeII ions
    with {\bvpc wavelengths (in Angstrom)} log$\,\lambda<\sim2.95$ and
    log$\,\lambda<\sim3.1$, in the {\it left} uppermost panel, for
    log$\,\lambda<\sim2.9$, the white region is formed as a result of negative
    radiative flux, which may be related to a drop in the source
    function \citep[\eg,][]{owocki99}.  We tested the influence of the negative fluxes on the
    total radiative acceleration by setting their values to zero at
    the place where radiative force is computed. The total radiative
    acceleration did not change and therefore the impact of those
    fluxes on the radiative acceleration should be small.  

  In order to understand which frequencies are important, in
  Fig.~\ref{fig:ACC_3D} (fifth panel from the top) we also display the
  contribution of \FeII ({\it left}) and \FeIII ({\it right}) to \Q of
  the respective ions (blue lines). These contributions are normalised
  in such a way that the sum over all frequencies would equal
  unity. The red line (with ordinate in red colour placed on the {\it
    right-hand side}) shows the sum of the contributions of \FeII (or
  \FeIII in the right panel) located in different frequency bins. It
  is evident that about 50\% of the acceleration of \FeII comes from
  lines with $\lambda$ between 2\,300 and 2\,800\,\AA, and the lines
  in the Balmer continuum provide more than 95\% of total acceleration
  of \FeII.

    \begin{table*}    
      \caption{Mass-loss rates at which $\Q=1$ for different model series.} 
      \begin{center}
		\tabcolsep1.2mm 
		\begin{tabular}{c c c l c c c c |c c| c c} % centered columns (6 columns)      
		  \noalign{\smallskip}
		  \hline		  
		  \noalign{\smallskip}
		  \multicolumn{8}{c}{$f_{\infty}=0.1;$ $\varv_{\rm cl}=30\,\kms$}  & \multicolumn{2}{|l}{\hspace{0.8cm}$Z=Z_\odot$}  & \multicolumn{2}{|c}{$Z=Z_\odot/2$} \\ %& \multicolumn{2}{|c}{$Z=Z_\odot/5$} & \multicolumn{2}{|c}{$Z=Z_\odot/10$} \\
		  log\,$\frac{\Lstar}{\Lsun}$ & ${\rm M}_\star$ &  $\Gamma_e$ & \mdot range  & $T_{\rm eff}$ & $R_\star$  & $\frac{\vinf}{\vesc}$ & \logg &  $\Gamma^{*}_{(\tau=2/3)}$& $\dot{M}_{\Q=1}$  & $\Gamma^{*}_{(\tau=2/3)}$ & $\dot{M}_{\Q=1}$ \\ 	 
        series name & ${\rm M}_\odot$ & & $10^{-6}{\rm M}_\odot/{\rm yr}$  & K           &$R_\odot$ &   & & &$10^{-6}{\rm M}_\odot/{\rm yr}$  & & $10^{-6}{\rm M}_\odot/{\rm yr}$  \\ 
		  \hline\noalign{\smallskip}
	      5.75  & 71  & 0.20  & $0.25-2.00$ & 30\,000 & \ 28 & 2.6 &	 3.40 & $0.70$ & 0.67 & $0.67$ & 0.48 \\
                    &     &       & $0.25-2.00$ & 27\,500 & \ 33 & 2.6 &	 3.25 & $0.67$ & 0.51 & $0.63$ & 0.39 \\ 
L5.75M71	    &     &       & $0.10-2.00$ & 25\,000 & \ 40 & 2.6 &	 3.09 & $0.62$ & 0.38 & $0.58$ & 0.31 \\ 
		    %\hline
		    &     &       & $0.10-2.00$ & 22\,500 & \ 49 & 2.6 &	 2.90 & $0.58$ & 0.30 & $0.55$ & 0.24 \\ 
		    &     &       & $0.25-2.00$ & 20\,000 & \ 62 & 1.3 &	 2.70 & $0.60$ & 1.80 & $0.57$ & 0.97 \\ 
		    &     &       & $0.25-2.00$ & 17\,500 & \ 81 & 1.3 &	 2.47 & $0.61$ & 1.55 & $0.57$ & 0.75 \\  
		    &     &       & $0.25-2.00$ & 15\,000 &  111 & 1.3 &	 2.20 & $0.58$ & 0.68 & $0.55$ & 0.42 \\ 
		    %\hline
		    &     &       & $0.10-2.00$ & 12\,500 &  159 & 1.3 &	 1.88 & $0.57$ & 0.28 & $0.55$ & 0.22 \\ 
		    &     &       & $0.30-1.00$ & 10\,000 &  249 & 1.3 &	 1.50 & $0.63$ & 0.10 &   --   &  --  \\
		    &     &       & $0.48-1.50$ &\ 8\,800 &  --  & 0.7 &	 1.27 & $0.75$ & 0.48 &   --   &  --  \\ 	
		  \hline\noalign{\smallskip}                                                              
	      5.50  & 40  & 0.20  & $0.10-2.00$ & 30\,000 & \ 21 & 2.6 &	 3.40 & $0.71$ & 0.39 & $0.67$ & 0.28 \\
                    &     &       & $0.10-2.00$ & 27\,500 & \ 25 & 2.6 &	 3.25 & $0.68$ & 0.29 & $0.63$ & 0.23 \\
L5.5M40		    &     &       & $0.10-2.00$ & 25\,000 & \ 30 & 2.6 &	 3.09 & $0.62$ & 0.22 & $0.58$ & 0.18 \\
                   %\hline
		    &     &       & $0.10-2.00$ & 22\,500 & \ 37 & 2.6 &	 2.90 & $0.59$ & 0.17 & $0.56$ & 0.14 \\
		    &     &       & $0.10-2.00$ & 20\,000 & \ 46 & 1.3 &	 2.70 & $0.61$ & 1.14 & $0.57$ & 0.58 \\ 
		    &     &       & $0.10-2.00$ & 17\,500 & \ 61 & 1.3 &	 2.47 & $0.61$ & 0.91 & $0.57$ & 0.43 \\ 
		    &     &       & $0.10-2.00$ & 15\,000 & \ 83 & 1.3 &	 2.20 & $0.57$ & 0.38 & $0.55$ & 0.24 \\
		   %\hline
		    &     &       & $0.10-2.00$ & 12\,500 &  120 & 1.3 &	 1.88 & $0.57$ & 0.17 & $0.55$ & 0.13 \\ 
		    &     &       & $0.04-1.00$ & 10\,000 &  187 & 1.3 &	 1.50 & $0.63$ & 0.06 & $0.62$ & 0.05 \\
		    &     &       & $0.25-1.50$ &\ 8\,800 &  242 & 0.7 &	 1.27 & $0.77$ & 0.53 &   --   &  --  \\ 
		  \hline                                                                                         
		  \noalign{\smallskip}		                                                                 
	      5.50  & 30  & 0.26  & $0.10-2.00$ & 30\,000 & \ 21 & 2.6 &	 3.28 & --     &  --  & $0.75$ & 0.36 \\
		    &     &       & $0.10-2.00$ & 27\,500 & \ 25 & 2.6 &	 3.13 & --     &  --  & $0.73$ & 0.28 \\
L5.5M30		    &     &       & $0.10-2.00$ & 25\,000 & \ 30 & 2.6 &	 2.96 & --     &  --  & $0.68$ & 0.20 \\
		    &     &       & $0.10-2.00$ & 22\,500 & \ 37 & 2.6 &	 2.78 & --     &  --  & $0.64$ & 0.16 \\
		    &     &       & $0.10-2.00$ & 20\,000 & \ 46 & 1.3 &	 2.58 & --     &  --  & $0.65$ & 0.88 \\ 
		    &     &       & $0.10-2.00$ & 17\,500 & \ 61 & 1.3 &	 2.34 & --     &  --  & $0.67$ & 0.60 \\ 
		    &     &       & $0.10-2.00$ & 15\,000 & \ 83 & 1.3 &	 2.08 & --     &  --  & $0.63$ & 0.31 \\
		    &     &       & $0.10-2.00$ & 12\,500 &  120 & 1.3 &	 1.76 & --     &  --  & $0.63$ & 0.16 \\ 
		    &     &       & $0.06-2.00$ & 10\,000 &  187 & 1.3 &	 1.37 & $0.69$ & 0.07 & $0.68$ & 0.06 \\
		    &     &       & $0.20-2.00$ &\ 8\,800 &  242 & 0.7 &	 1.27 & $0.79$ & 1.43 & --     & --   \\ 
		  \hline                                                                                   
		  \noalign{\smallskip}			                                                   
              5.50  & 20  & 0.39  & $0.07-0.50$ & 10\,000 &  187 & 1.3 &	 1.19 & $0.79$ & 0.12 & $0.78$ & 0.09 \\
L5.5M20		    &     &       & $1.25-5.50$ &  8\,800 &  242 & 0.7 &	 0.97 & $0.84$ & 4.04 & $0.86$ & 1.85 \\	
		  \hline
		\end{tabular}
		\label{tab:mdot_pred} % is used to refer this table in the text
      \end{center}
      \small
      {\bf Notes.} The {\it physical} Eddington factor, $\Gamma$, is distance dependent and therefore we show the value of $\Gamma$ at the reference radius, where $\tau_{\rm ROSS}=2/3$.
    \end{table*}

  To understand the significance of these numbers, in the lowermost
  panels of the figure we present the contribution of \FeII (left)
  and \FeIII(right) to the {\it total} work ratio, provided by all
  ions. In the cooler model 40\% of the total acceleration arises from
  lines with $\lambda$ between 2\,300 and 2\,800\,\AA \, and the \FeII
  lines in the Balmer continuum provide about 70\% of the total
  acceleration. In the hotter model, the Balmer continuum also
  provides a significant fraction of the total radiative force (about
  50\%).
  
  According to Wien's displacement law, a black body with $\Teff=8\,800$\,K  peaks its radiative flux at $\lambda\sim3\,300$\,\AA, which is near the wavelength interval where most of total line acceleration of the cooler model is provided. This is expected to support the total line force along with the larger fraction of \FeII. However, with a further decrease of \Teff, despite the expected increase of the \FeII fraction, the overall radiation decreases, whilst furthermore the peak of radiation flux moves towards longer wavelengths. Thus, we do not expect that the size of the second bi-stability jump, if it were determined by models cooler than 8\,800\,K, to be significantly different from the one determined here between 10\,000 and 8\,800\,K.

    The theory of radiation-driven winds predicts that \mdot depends on metallicity in the following way:
    \begin{equation}
     \mdot\propto Z^m
    \end{equation}
    with
    \begin{equation}
     m = \frac{1-\alpha}{\alpha - \delta}
    \end{equation}
    where $\alpha$ quantifies the ratio of the line force from optically thick lines to the line acceleration of all lines. The radiative acceleration is caused by an assortment of optically thin and thick lines and therefore $\alpha$ is between 0 and unity. The parameter $\delta$ was introduced  by \citet{abbott82} to account for variation of the ionisation throughout the wind \citep[see also][]{kudritzki89}.    
    The typical values for $m$ are ranging from 0.5 \citep*{kudritzki87} to 0.94 \citep{abbott82}. \citet{vink01} found that $m=0.69$ for O stars and $m=0.64$ for B supergiants.

\begin{figure}
	%\centering 
	\hspace{0.65cm}\resizebox{0.8\hsize}{!}{\includegraphics{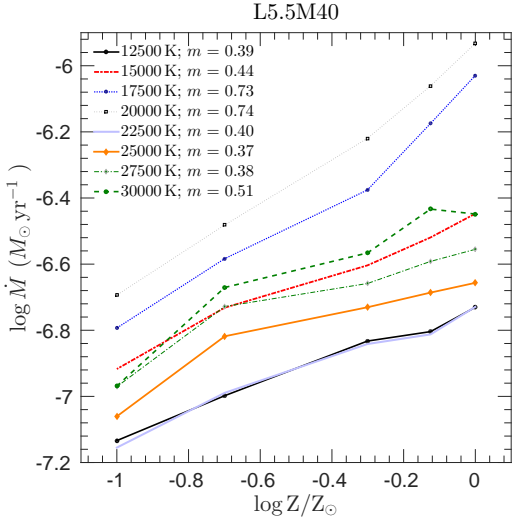}}
	\caption{Stellar wind mass loss as a function of metallicity for models with different temperatures. }	
	\label{fig:MdotvsZ}
    \end{figure}

    Figure~\ref{fig:MdotvsZ} shows that the value of $m$ strongly
    depends on \Teff.  \mdot displays a weaker dependence on
    metallicity in the models with $\Teff=12\,500$ and 22\,500\,K, as
    \FeIII is not an important line driver for these
    temperatures. Consequently, \mdot becomes less sensitive to the
    adopted metal abundances.  On the other hand, at $\Teff=20\,000$,
    \FeIII is the dominant wind driver, and therefore the radiative
    acceleration becomes sensitive to metal abundance. Note that
    between 22\,500 and 20\,000\,K $m$ increases from 0.4 to 0.74,
    which indicates that the winds are driven by different ions.

\section{Comparison with Monte-Carlo mass-loss rates}
\label{sec:comp}

   In Figure~\ref{fig:GALBJ_L5.5M40_ISO}, we compare our results with the mass-loss rates following from the recipe of \citet{vink00,vink01}, which are used both in observational work and in all up-to-date evolutionary models for massive stars. What complicates a meaningful comparison is the fact that the temperature location of the bi-stability jump predicted by \cmfgen and Monte-Carlo calculations is different. 
   {\bvpb
   This means that we can only make rough comparisons between the two different methods, as it is not very meaningful to compare the mass-loss rate provided by one code at one \Teff to the mass-loss rate at the same \Teff by the other code. The differences we discuss in the following are thus only indicative.   
   {\bvpc Due to the different temperature locations of the bi-stability jump in both methods, a prescribed \vinf/\vesc value (corresponding to an option of '$\vinf= -1$' in the script) in V00/V01 idl routine to
   calculate \mdot is not appropriate to use (especially for models near BSJ).}
   Instead, in the recipe we provide the absolute values of \vinf in \kms as used in our \cmfgen models.
   We expect such a comparison to provide a better flavor of the differences between \cmfgen and V00/V01 mass-loss rates.
   
   Because of the different temperature locations of the bi-stability jump in both codes, we compare our mass-loss rates to V00/V01 mass-loss rates for
   $\Teff\geq25\,000$\,K, \Teff between $20\,000$ and $15\,000$\,K, and for $\Teff<10\,000$\,K , \ie when in both methods the models are above, in between, or below the two bi-stability jumps.
   }      
   For L5.5M40 models series with solar metal abundance, we find the following (other cases lead to similar conclusions):   
   \begin{itemize}
    \item above the first bi-stability jump ($\Teff\geq25\,000$\,K), the mass-loss rates are similar
    \item \cmfgen predicts that the bi-stability jump should occur at lower temperature (between 22\,500 and 20\,000\,K) in comparison to the theoretical expectations of V00/V01 (between 25\,000\,K and 22\,500\,K), which is in good agreement to the observations.
    Interestingly, the models just above the first BSJ according to V00/V01 have similar rates to those just above the first BSJ in \cmfgen, \ie $\dot{M}^{25\,000}_{\rm vink}\approx\dot{M}^{22\,500}_{\rm \Q=1}$.

    \item over the first BSJ, in \cmfgen mass-loss rates are increased by a factor of 6 (4 for Z/Z$_\odot=0.5$), whilst 
    over the second BSJ, mass-loss rates are increased by a factor of about 10. In \cmfgen, the size of both BS jumps is strongly 
    dependent on the proximity of the models to the Eddington limit (\cf Fig.~\ref{fig:BJvsGamma}).
    On the other hand, according to V00/V01, both BS jumps should increase \mdot roughly by a factor of ten,
    independent of metal abundances and the Eddington factor. 
    
    \item %In general, mass-loss rates predicted by V00/V01  are larger than \cmfgen and  from hotter to cooler  temperatures, the discrepancies between \cmfgen and V00/V01 mass-loss rates become larger. 
    despite that our models are inhomogeneous ($\fcl=0.1$), 
    the recipe of V00/V01, which is based on homogeneous wind models, 
    predicts higher \mdot than  \cmfgen %and the discrepancies between  \cmfgen and V00/V01 mass-loss rates become larger with reducing the temperature
    \footnote{{\bvpb This is interesting because non-homogeneous models are expected to lead to higher mass-loss rates as the presence of clumping increases density inside the clumps which in turn increases the recombination rates. Consequently, the winds become less ionised. As lower ions have larger number of driving lines than the higher ones, the radiative force should increase \citep{muijres11}. 
    \cite{sundqvist14} concluded that porosity in velocity space typically gives higher {\it empirical} mass-loss rates, but also a downward correction in theoretical mass-loss rates is possible if significant velocity-porosity at the wind critical point is present.
    }}. 
      Overall, the mass-loss rates in the cool Bsg range (between $20\,000$ and $15\,000$\,K) are a factor of 2-5 lower than V00/V01 mass-loss rates. 
      {\bvpb A possible reason for the lower mass-loss rates
      predicted by \cmfgen is that we are not sure whether we use all important driving
      lines, whilst \cite{vink00,vink01} pre-selected $10^5$ relevant driving lines from millions bound-bound transitions for the first 30 elements in the Periodic Table.
      }
      %that we use only a subset of the most important
      %ions for the line driving (\cf Table~\ref{tab:atm_data}), whilst \citet{vink01} used {\bvpb millions}
      %line transitions of the first 30 elements from the periodic table from H to Zn.
      Furthermore, \cite{vink00,vink01} calculations are based on absolute solar metal abundances, $Z_\odot=0.019$, taken from       \citet{anders89} which are larger than those determined by \citet{asplund09}, used in this investigation.
    \item below the second BSJ,
    at $\Teff=8\,800$\,K, V00/V01 rates are a factors of 6 to 40 larger than \cmfgen mass-loss rates. The largest discrepancy between Monte-Carlo predictions and \cmfgen is at \Teff=10\,000\,K, where V00/V01 mass-loss rates are a factor of 60 to 85 larger (\cf Fig.~\ref{fig:GALBJ_L5.5M40_ISO}). 
   Such a large discrepancy may be attributed to the different temperature locations of the second BSJ in \cmfgen and Monte-Carlo calculations. 
   As an example, in the L5.5M40 model series the second BSJ is located between 10\,000 and 8\,800\,K, whilst V00/V01 rates predict a second BSJ between 15\,000 and 12\,500\,K. Consequently, at $\Teff=10\,000$\,K, V00/V01 mass-loss rates are increased by the second BSJ, whilst in \cmfgen a second BSJ has not yet happened.
  \end{itemize}
  {\bvpb
  One should note that even in the absence of any bi-stability jumps (first or second) the overall mass-loss rate is expected to decrease with dropping \Teff. Therefore, if one were to compare a hotter Monte-Carlo model with a cooler \cmfgen model, we would anticipate the Monte-Carlo model to have a higher \mdot than the rate determined by the \cmfgen model, simply due to their different \Teff.
  One should also be aware that most stellar evolution modellers 'switch' from \cite{vink99,vink00,vink01} to \cite{Nieuwenhuijzen90} for cooler models. The relevance of this is that if we now find that the second jump takes place at lower \Teff than in the V00/V01 recipe, then the mass-loss rate in the stellar evolution models would be lower than assumed {\sc for those temperatures}.  It remains to be investigated whether the overall mass lost would be affected significantly, as the dramatic increase may or may not fully take place at a lower \Teff.  
  %However, the overall mass lost would not necessarily be affected, as the dramatic increase would now just take place at a lower \Teff.
  }

\section{The bi-stability jump and Luminous Blue Variables}
  \label{sec:LBVs}
 
    The bi-stability jumps, discussed in the previous section in the
    context of normal OB supergiants, might also play a role in the
    mass loss behaviour of LBVs. 
    Based on the bi-stability mechanism, \citet{vink02} and \citet*{groh11a} 
    were able to
    partly explain the behaviour of the \mdot found by \citet{stahl01}
    during a typical LBV variation of AG\,CAR. However, in their
    calculations for LBV winds, the second jump occurred at a significantly 
    hotter temperature ($\Teff\simeq20\,\,000$\,K) in comparison to
    observations, whilst our \cmfgen calculations suggest that
    the second jump should occur at a cooler temperature
    ($\Teff\simeq10\,\,000$\,K).
    
    As the nature of LBVs is still under debate,
    their variable winds might offer key prospects to understanding their role as direct progenitors of SNe because 
    both LBVs and SNe have shown double-troughed \Ha line
    profiles, which can be explained if their wind changes
    instantaneously - as expected from the bi-stability jump \citep{trundle08,groh11}.
  
    Due to their high mass-loss rates LBVs have lost a considerable amount of mass 
    during their
    evolution. 
    Their luminosity-to-mass ratio is higher than for normal AB supergiants, and 
    they might thus be in critical proximity to their Eddington limit \citep{hd94,vink12}.
    It may thus be valuable to understand
    the influence of the Eddington factor on the size of the
    bi-stability jump. 

\begin{figure}
	\centering 
	\resizebox{0.8\hsize}{!}{\includegraphics{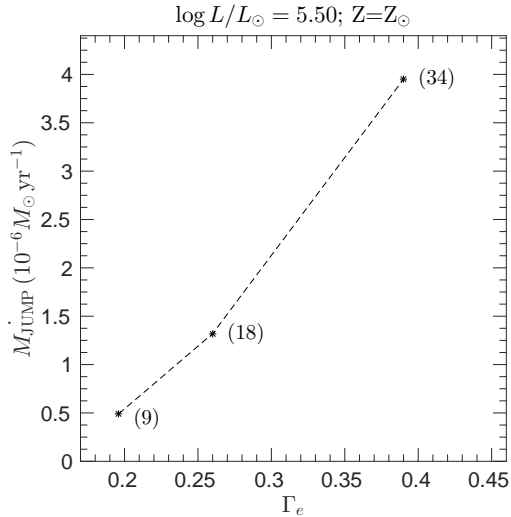}}
	\caption{Dependence of the jump in \mdot  between 10\,000 and 8\,800\,K on Eddington factor. The numbers in parentheses show the increase of mass loss rate across the second bi-stability jump in relative sense, \ie $\dot{M}_{\Q=1}^{8\,800}/\dot{M}_{\Q=1}^{10\,000}$.}	
	\label{fig:BJvsGamma}
    \end{figure}

    In Fig.~\ref{fig:BJvsGamma}, the size of the
    jump between $10\,000$ and $8\,800$\,K as a function of {\it
    classical} Eddington factor is illustrated. 
    An increase of $\Gamma_e$ leads to an increase in \mj, which is in agreement with the results  of \citet{vink02} for LBVs and \citet{vink06}, \citet{grafener08} and \citet{vink11} for  very massive stars (VMS) as Of/WNh type objects.     
      {\bvpb Across the second bi-stability jump mass-loss rates are increased by a factor of between 10 and 30.} This is caused by two effects: i) an increase of \mdot due to the second BSJ; and ii) an increase of \mdot due to the Eddington limit.    
    %An increase of \mj
    %with $\sim1-4\times10^{-6}\msunpyr$ across the second bi-stability
    %jump is significant and, 
    If such an increase of \mj is real, we may expect to find
    different wind properties of objects located on both sides of this
    second jump, which would be particularly strong for objects close to the Eddington limit.

\section{Conclusions}
\label{sec:conclusions}

  We have calculated a grid of wind models and predicted mass-loss
  rates of BA supergiants using the \cmfgen code.
  Our calculations independently confirm the bi-stability jump in \mdot around $\Teff\simeq22\,500$\,K predicted by \citet{vink99}. However, \cmfgen predicts that this jump will occur at a lower temperature $\Teff\simeq20\,000$\,K ($\Teff=17\,500$\,K if clumping is not taken into account), which is consistent with observations. 
    
  In our models, the ions of C, N, and, O are the most important line
  ``drivers`` for $\Teff>22\,500$\,K, whilst for temperatures between
  20\,000 and 12\,500\,K \FeIII provides most of the radiative
  acceleration. For temperatures below 10\,000\,K, \FeIII starts to
  recombine to \FeII and at $\Teff=8\,800$\,K we find that \FeII
  provides most of the total line acceleration. This causes a second
  bi-stability jump in \mdot around 8\,800\,K which is {\it dependent} on
  the Eddington parameter and metallicity. 
  However, in many cases the V00/V01 recipe predicts a jump in \mdot by a factor of about 10, {\it independent} of metal abundance or proximity to the Eddington limit.
  {\bvpb
  According to \cmfgen, this second jump also takes place at cooler temperature than the predicted temperature of the jump by V00/V01.  
  
  In the context of stellar evolution not only the discrepancies regarding the temperature position of the jump are important, but also the  discrepancies in the absolute mass-loss rates.
  Most evolutionary models of massive stars use mass-loss rates from  V00/V01 recipe. Consequently, the reported discrepancies between \cmfgen and Monte-Carlo predictions regarding the second bi-stability jump,
  %might provide inputs in our understanding of massive star evolution, which can help to better understand the relation between different types of massive stars   
  imply that the mass-loss rates in late B and A-supergiants are likely too high, and need to be considered 
  with caution when used in evolutionary calculations.
  Regarding the first jump the discrepancies between both methods are much smaller and the mass-loss rates are similar.
  }
  
  We found that at half-solar metal abundances a second BSJ is produced only in models close to  the Eddington limit (with $\Gamma_e\sim0.39$), whilst  for normal BA supergiants the second BSJ is produced only for solar metal composition. Thus, for LBVs the second BSJ should be important even in low-metallicity environments and  can be used as a tool to better understand the observed variations
  in their mass-loss rates. As the nature of LBVs
  is still not well understood, a detailed investigation of the second BSJ would be valuable.

  {\bvp A relevant question is whether the by \cmfgen predicted BS
  jumps in \mdot are artificially produced due to the applied velocity ratios?
  We have demonstrated that for constant 
  velocity ratio, \cmfgen predicts a jump in \mdot between $22\,500$ and
  20\,000\,K (Fig.~\ref{fig:BJ_L5.5M40}), thus it is likely that the
  observed velocities are a consequence of an increase/drop in
  \mdot around 21\,000\,K. However, this does not imply that the mass-loss rate 
  varies with the same factor as the velocity ratios, because the
  reactions can be non-linear, due to the non-linear acceleration and the different
  elements which initiate and accelerate the wind.  
  On the other hand, the second BSJ is only produced in models in close proximity to Eddington limit
  when the velocity ratio is kept fixed. This underlines the importance of both bi-stability jumps for LBVs. 
  It is relevant to mention that the exact temperature of the bi-stability
  jumps remains somewhat ambiguous, as the ionisation equilibrium of Fe
  is sensitive to  density and clumping properties in the lower wind and thus to  stellar luminosity and mass.
  
  Knowledge of the effects of clumping on the second bi-stability
  jump might be valuable, especially for LBVs, as they experience
  outbursts and episodes of enhanced mass loss during which the degree
  of clumping might change. Thus, the driving efficiency of Fe-group
  elements might be different for specific temperature at various
  epochs during LBV phase. Therefore, it is important to quantify the effects of
  clumping on both sides of both bi-stability jumps \citep[see][]{davies07}. 
  Understanding all that, we
  might be able to explain some of the observed variations in \mdot
  during the different phases of LBVs, with relevance for the properties of SN types IIb and IIn.
}
 
 \section*{Acknowledgements}

    We acknowledge financial support from the Northern Ireland Department of Culture, Arts, and Leisure (DCAL) and the United Kingdom (UK) Science and Technologies Facilities Council (STFC).
    We gratefully thank Dr. John Hillier for providing the \cmfgen code  to the astronomical community and Dr. Joachim Puls for his useful comments towards the improvement of the paper.

\bibliographystyle{mn2e}
\bibliography{bpbib}

\begin{appendix}
  %\vspace{10.5cm}
  \vspace{1.2cm}
  \section{Model atom (sophisticated models)}
  
  Adopted atomic data of all elements included in our model atmosphere calculations are summarised in Table~\ref{tab:atm_data}. The main source of atomic data  comes from the Opacity project \citep[][]{seaton87} and the Iron project \citet{hummer93}. However, for some CNO elements, atomic data were used also from \citet{nussbaumer83,nussbaumer84}, whilst for \FeII, \FeIII, \FeIV, and \FeVII data were used also  from \citet{nahar95,zhang96,becker95,becker95b} respectively.
  
  In order to save computational time, at the different temperature regimes we choose, different (but relevant) level assignments for the involved ions. 
  
  \begin{table}
    \caption{Atomic data included in our models}
    \label{tab:advanced_models}
    \begin{center}
      \begin{tabular}{l| r r r r r } % centered columns (4 columns)
      \hline  
      \Teff range [kK]& 8.8-10    & 12.5-20 & 22.5-27.5 & 30 \,\,\,\,       &\\
       Ion \\
      \hline
      \HI       & 20/30    & 20/30   & 20/ 30    & 20/ 30    &\\  
      \HeI      & 45/69    & 45/69   & 45/ 69    & 45/ 69    &\\  
      \HeII     & 22/30    & 22/30   & 22/ 30    & 22/ 30    &\\    

      \CI       & 22/42    &  --     &  --       &  --       &\\
      \CII      & 104/338  & 104/338 & 31/68     & 10,10,18  &\\
      \CIII     & 91/209   & 91/209  & 99/243    & 99/243    &\\  
      \CIV      & --       & 19/24   & 59/64     & 59/64     &\\
      \CV       & --       &  --     &  --       & 46/73     &\\
      \NI       & 22/35    &  --     &  --       &  --       &\\
      \NII      & 100/267  & 100/267 & 80/192    & 9/17      &\\  
      \NIII     & --       & 41/82   & 41/82     & 41/82     &\\
      \NIV      & --       & 13/23   & 78/124    & 200/278   &\\  
      \NV       & --       &  --     &  --       &  --       &\\  
      \OI       & 32/161   & 13/29   &  --       &  --       &\\
      \OII      & 137/340  & 137/340 & 106/251   &  81/182   &\\ 
      \OIII     &  --      & 165/343 & 165/343   & 165/343   &\\ 
      \OIV      &  --      & 9/16    & 99/202    & 71/138    &\\ 
      \OV       &  --      &  --     &  --       & 11/19     &\\ 
      %O  VI     &  --      &  --     &  --       &  --       &\\ 
   \ion{Ne}{ii} & 42/242   & 42/242  & 42/242    & 25/116    &\\ 
   \ion{Ne}{iii}&  --      & 20/51   & 57/188    & 57/188    &\\ 
      %Ne IV     &  --      & --      &  --       &  --       &\\ 
   \ion{Mg}{ii} & 22/65    & 18/36   &  --       &  --       &\\ 
   \ion{Mg}{iii}& 41/201   & 41/201  & 41/201    & 41/201    &\\ 
%      AL I      &  --      & ---     &  --       &  --       &\\  
   \ion{AL}{ii} & 37/56    & 37/56   &  --       &  --       &\\ 
   \ion{AL}{iii}& 18/50    & 18/50   & 18/50     & 7/12      &\\ 
   \ion{AL}{iv} &  --      & 46/107  & 46/107    & 62/199    &\\
   \ion{Si}{ii} & 27/53    & 27/53   &  --       &  --       &\\ 
   \ion{Si}{iii}& 81/147   & 81/147  & 81/147    & 26/51     &\\  
   \ion{Si}{iv} &  --      & 39/50   & 55/66     & 55/66     &\\ 
   \ion{Si}{v}  &  --      & 12/22   & 52/203    & 52/203    &\\ 
      \PIV      & 30/90    & 30/90   & 30/90     & 30/90     &\\ 
      \PV       &  --      & 9/15    & 9/15      & 9/15      &\\  
   \ion{S}{ii}  & 41/171   & 41/171  & 12/33     &  --       &\\ 
   \ion{S}{iii} & 80/257   & 80/257  & 80/257    & 41/83     &\\ 
   \ion{S}{iv}  &  --      & 49/138  & 69/194    & 69/194    &\\ 
   \ion{S}{ii}  &          & 9/15    & 17/36     & 41/167    &\\
   \ion{Ar}{iii}& 29/249   & 29/249  & 29/249    & 18/82     &\\ 
   \ion{Ar}{iv} &  --      & 8/22    & 29/97     & 41/204    &\\  
%      AR V      &  --      & --      &  --       &  --       &\\  
   \ion{Ca}{ii} & 21/70    & --      &  --       &  --       &\\ 
   \ion{Ca}{iii}& 41/208   & 41/208  & 41/208    & 41/208    &\\ 
   \ion{Ca}{iv} &  --      & 2/3     & 33/171    & 39/341    &\\ 
%      Ca V      &  --      &  --     &  --       &  --       &\\ 
%      Fe I      &  --      &  --     &  --       &  --       &\\  
      \FeII     & 275/827  & 17/218  &  --       &  --       &\\ 
      \FeIII    & 136/1500 & 136/1500& 136/1500  &  --       &\\ 
      \FeIV     &  --      & 74/540  & 100/1000  & 100/1000  &\\ 
      \FeV      &  --      & 17/67   & 34/352    & 45/869    &\\ 
      \FeVI     &  --      &  --     &  --       & 55/674    &\\ 
      \FeVII    &  --      &  --     &  --       & 13/50     &\\
     \hline
     \end{tabular} 
   \end{center}    
   \small
     {\bf Notes.} For each ion, the number of super levels and full levels are provided.
   \label{tab:atm_data} % is used to refer this table in the text
\end{table}

  \end{appendix}

\end{document}